**Fitting, Evaluating, and Comparing Cognitive Architecture Models Using Likelihood:**

**A Primer With Examples in ACT-R**

Andrea Stocco[1], Konstantinos Mitsopoulos[2], Yuxue C. Yang[1,3], Holly S. Hake[1,4],

Theodros Haile[1,4], Bridget Leonard[1], and Kevin Gluck[2]

[1] Department of Psychology, University of Washington, Seattle, WA 98195

[2] Institute for Human and Machine Cognition, Pensacola, FL

[3] Now at Activision, Los Angeles, CA

[4] Now at the University of Groningen, The Netherlands

Correspondence To:

Andrea Stocco,

Department of Psychology,

Campus Box 351525

University of Washington,

Seattle, WA 98195

Email: stocco@uw.edu

Phone: +1 206-543-4159

**Abstract**

Cognitive architectures are influential, integrated computational frameworks for modeling cognitive processes. Due to a variety of factors, however, researchers using cognitive architectures to explain and predict human performance rarely employ model validation, comparison, and selection techniques based on likelihood. This paper provides a primer on how to implement maximum likelihood techniques and its derivatives to fit and compare models at the individual and group level, using models implemented in the ACT-R cognitive architecture as examples. The paper covers the most common ways in which likelihood measures can be applied, under different scenarios, for models of different complexity, and provides further technical references for the interested reader. An accompanying notebook in Python provides the code to implement all of the suggestions.

**Introduction**

Cognitive architectures are integrated computational frameworks that are often used to model human behavior in cognitive psychology and cognitive neuroscience [1,2]. Rooted in ideas first promoted by Allen Newell [3], cognitive architectures stress the importance of modeling cognition in an integrated manner, including the combined contributions of perceptual, control, and memory systems.

As part of their work, cognitive architecture researchers often perform operations of model fitting, attempting to maximize the goodness-of-fit of the model parameters against some group-level (or, more rarely, individual-level) behavioral measure(s). The measures used for model fitting fall into two broad categories: measures of *closeness* and measures of *likelihood* [4]. At the time of writing, the most common measures used to fit these models are closeness measures, such as Spearman correlations ($r$) and Mean Squared Error (MSE). For example, out of 15 ACT-R journal papers published in the past two years, none makes use of likelihood to evaluate models.

In this paper, we argue that this approach is out of step with the standards currently adopted in other computational fields, which instead rely on *likelihood*. Likelihood is a foundational concept in statistics, and has been advocated for and used to estimate parameters and compare formal models in mathematical psychology [5], computational neuroscience [6], computational psychiatry [7], and computational biology [8].

By adopting likelihood, researchers in the cognitive architecture domains can take advantage of modern ideas in statistics and make deeper connections with other modeling communities. This paper is designed to facilitate this process by introducing a reader who is already familiar with cognitive architectures to the use of likelihood in their research, explaining why and where it is advantageous, and outlining how it can be used at different moments of the model development and evaluation pipeline.

**What is Likelihood?**

Likelihood is a statistical concept that connects a model and its parameters to the observable data. Formally, the likelihood of a model $m$ with a given set of parameter values $\theta$, given the data $\mathbf{x}=\mathbf{x}_0$, is the value of the probability distribution that generated the data, evaluated at the observed data $\mathbf{x}_o$ [5,9]:

$$\mathscr{L}(m, \theta \mid \mathbf{x}_o) = P(\mathbf{x} = \mathbf{x}_o \mid m, \theta) \tag{1}$$

The notation $P(\mathbf{x} = \mathbf{x}_o \mid \cdot)$ denotes that the likelihood has the same functional form as the probability distribution that generated the data, and it defines a function in the parameter space given the observed data. Likelihood provides a comparable quantitative assessment of how plausible the observed data are under specific parameter values.

A subtle but important distinction to make is that, although every likelihood value can have a probabilistic interpretation, the likelihood function itself is *not* a probability distribution, and likelihood is not a probability. That is because the likelihood and probability functions in Equation 1 have different domains: likelihood $\mathscr{L}$ is defined over the domain of *parameter values* for a given model, while the probability $P$ is a distribution over the domain of possible *data*. Thus, given a model and its parameters, it is possible to define a probability distribution over the possible observable results $\mathbf{x}$, and values of this distribution will indeed have the characteristics of probabilities—for example, the integral of the distribution will be one). The likelihood values over different parameters, however, do not behave the same way: they are not probabilities and thus do not need to integrate to one. Figure 1 illustrates the relationships and differences between the likelihood function of a model over its parameter values and the probability densities of the data, given the parameters.

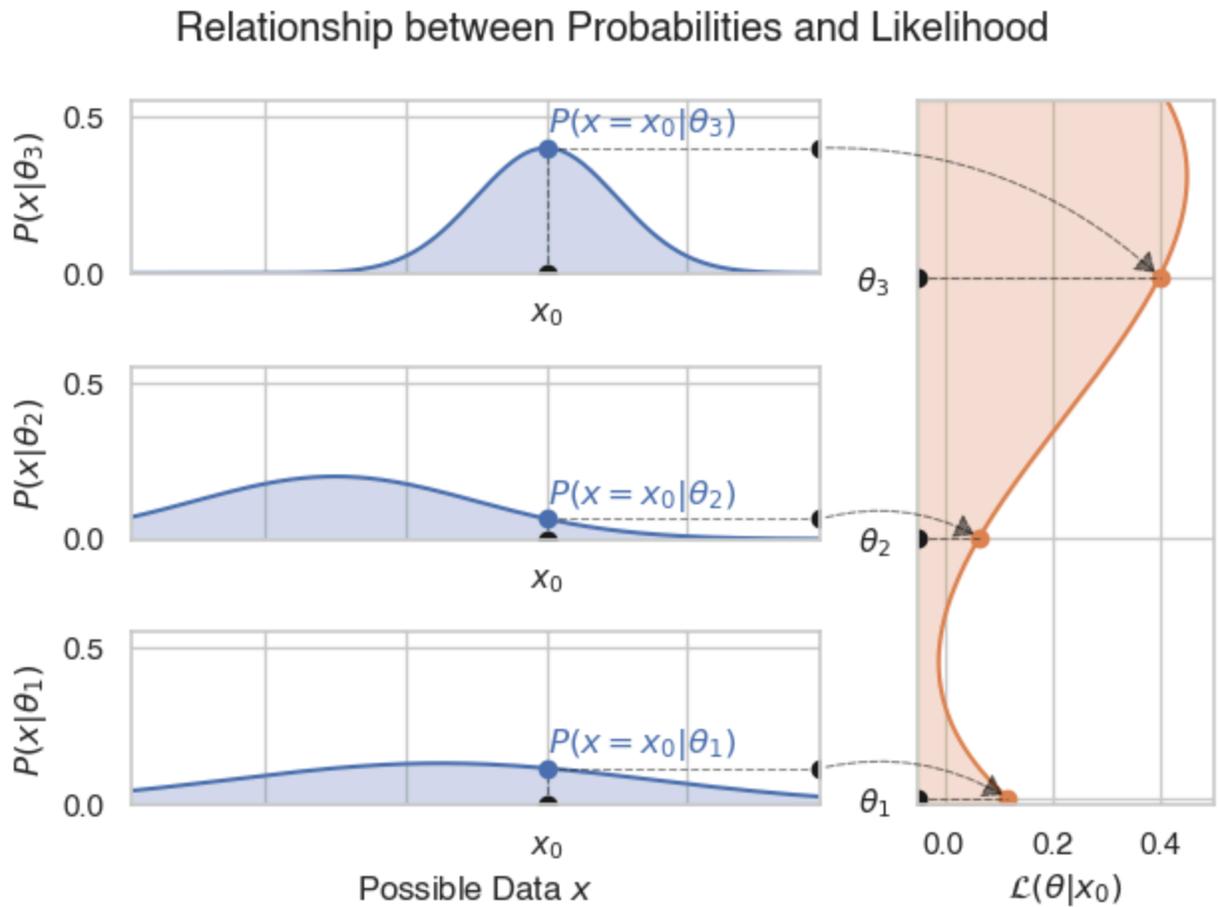

**Figure 1**: Difference and relationship between likelihood and probability functions. Likelihood functions are defined over the values of parameters **θ** of a model; for every value of the parameters, the corresponding likelihood value is the value of the probability density (or mass) function that describes how the model would generate the observed data **x₀**, conditioned on these parameters.

**Why Should You Use Likelihood?**

There are at least three reasons why a researcher would prefer likelihood over closeness measures of goodness of fit. The first is that likelihood values, as implied by Equation 1, represent how plausible the observed data are under the model's parameters. This is in contrast to the other commonly used metrics for comparisons, such as the mean squared error (MSE), whose interpretation is opaque and strictly measure-dependent. Furthermore, because it can be connected to the value of a probability density function *over the data*, it does not vary with respect to the units of measurements that are used to record the data. Likelihoods are also naturally expressed on the same scale for all possible measures, allowing combinations of disparate measures, for example, response times and accuracies—something that would be impossible to do with correlation coefficients or MSEs.

The second reason to use likelihood is that it is an important tool in *model fitting*, that is, the search of parameters $\boldsymbol{\theta}^*$ that maximize the fit of a model *m* to the data **x**. Instead of minimizing an error function, such as MSE, researchers can instead search for parameters that maximize their model's likelihood:

$$\boldsymbol{\theta}^* = \text{argmax}_{\boldsymbol{\theta}} \, P(\mathbf{x} = \mathbf{x}_o \mid m, \boldsymbol{\theta}) \tag{2}$$

This is, of course, the Maximum Likelihood Estimation (MLE) principle, which also underlies most statistical analysis, including the General Linear Model. It is also an important mechanism in *inference* and *model-based assessment* of individuals [10]: given some observable data, it is possible to identify the most likely parameter values that would produce that data, and use these values as measurements of some latent construct from the data.

Advocating for the use of likelihood instead of common error measures might seem a purely academic distinction—wouldn't reducing the error of a model be equivalent to maximizing its likelihood? In general, the answer is *no*. They happen to be equivalent for the specific case of *linear* models with *normally distributed* errors: when fitting a regression line, maximizing the likelihood and minimizing the sum of squared errors are, indeed, equivalent (and both equivalent to maximizing the correlation between observed and predicted data). This equivalence does not hold for *non*-linear models or when errors are *non*-normally distributed. Consider, for example, the case of the four data points in Figure 2. The points are generated by an exponential function $y = ae^{x-b}$, plus some measurement error. The measurement error is proportional to the value of $x$: the bigger $x$ is, the greater is the error that is associated with the measure. As a result, the fourth, rightmost point in the figure does not line up with the line that would best fit the other three.

In this scenario, we want to find the parameter values for $a$ and $b$ that best fit the data. The orange line is the solution found by minimizing the mean squared error (MSE) between predicted and actual data. The green line, instead, is the solution found by maximizing likelihood. When using MSE, all points are treated equally: what matters, in selecting the values of $a$ and $b$, is only the difference between the predicted and observed points. Because of this, the difference between the value of the fourth point and the continuation of the first three has a large impact on the estimated parameters. When maximizing likelihood, however, we can take into account that fact that points to the right are associated with greater error and, therefore, might still be plausibly generated by the model despite their large prediction error. As a result, fitting based on likelihood results in greater attention paid to the error of the three points to the left.

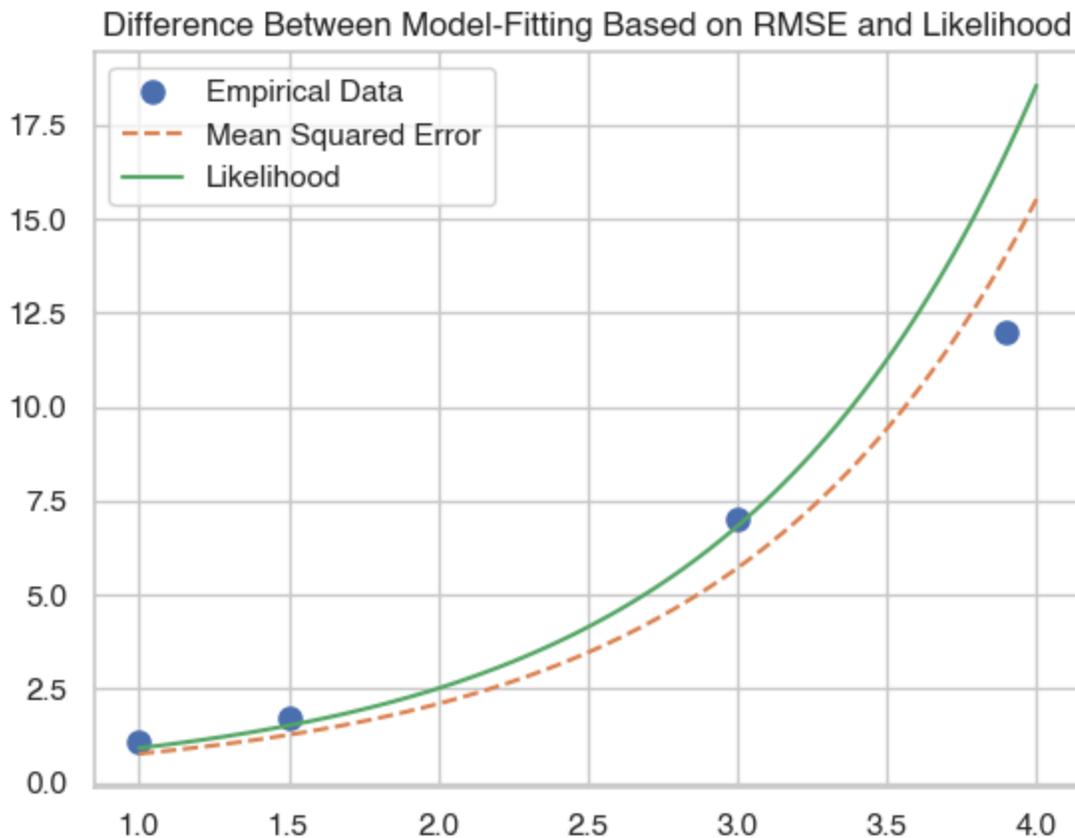

**Figure 2**: Fitting a set of data (the four blue points) using the same non-linear model (a function of the form $y = ae^{x-b}$) produces different results when minimizing the mean squared error vs. maximizing the likelihood of producing the observed data.

Because the core equations underlying ACT-R and other cognitive architectures are often non-linear, error-based measures like MSE are biased and more likely to be thrown off by observations that are associated with greater errors (for example, outliers). In turn, this leads to using parameters that are not truly representative of the underlying data.

The identification of best-fitting and interpretable parameters is particularly important when these parameters are used for *idiographic* model fitting—that is, models parameterized for a single individual rather than for a group average. As the rest of the paper will show, likelihood is straightforward to apply at the individual level, and it is easy to extend the individual model likelihoods to a group-level likelihood. This flexibility is particularly important because modeling individual differences has become an important application of computational models [10–14]. This trend is reflected in the application of cognitive architecture to model complex behavior of individual participants [15–24] and their application as "cognitive twins" or "digital twins" [25–27].

Finally, likelihood plays an important role in model comparison and selection. The two most common criteria, Akaike's Information Criterion (AIC: [28]) and Schwarz's Bayesian Information Criterion (BIC: [29]), were derived and explicitly defined in terms of $\mathscr{L}$ As this paper will later demonstrate, using likelihood makes the use of AIC and BIC in comparing ACT-R models trivial.

**Why Isn't Likelihood Commonly Used in Cognitive Architectures?**

There are practical and historical reasons why likelihood has not been commonly used in cognitive architectures. Cognitive architectures, by their very nature, are complex modeling paradigms; this makes them harder to analyze mathematically and simulate computationally. Because of their integrated nature, they are often applied to model human behavior in complex tasks, which are also intrinsically harder to analyze.

The main obstacle to this is that there are no closed-form equations that transform data values into the most likely parameters for ACT-R models—something akin to the algebraic solutions that exist for linear models [4]. In fact, while cognitive architectures do provide sets of equations that govern their inner workings, they do not typically provide even the exact probability distributions for the outcomes of said equations.

In the simplest cases, however, these probabilities can be calculated exactly and directly, even on a trial-by-trial basis, in a single run of the model. And, in the specific case of ACT-R, Fisher, Houpt, and Gunzelmann [30,31] have done remarkable work deriving the analytical expressions of the probability distributions of all of its critical component equations—and even for the tutorial ACT-R models that are included with the software.

But in *every* case, no matter how complex the model, likelihoods can be computed empirically using repeated runs of the model. After the empirical data distributions are known, models can be fitted using modern optimization techniques that do not require explicit derivatives of likelihood functions and drastically reduce the number of simulations needed. All of these cases will be covered in the following sections.

**How This Paper is Structured**

This paper provides a primer on how to fit, evaluate, and compare models developed within cognitive architectures on the basis of their likelihood. The examples provided herein are implemented in one specific cognitive architecture, known as *Adaptive Control of Thought — Rational* or simply ACT-R. This choice was made for the practical reason that the authors have experience with modeling in ACT-R and for the tactical reason that ACT-R is currently the most widely used and most broadly validated architecture within the psychological sciences [1], having approximately 1,000 scientific publications to its name. In addition to its popularity, ACT-R is representative of cognitive architectures in general, sharing fundamental assumptions that are common to other influential architectures, such as Soar [32] or SIGMA [33] and that arguably can be found in the architecture of the human brain [34]. Thus, examples provided in the context of ACT-R should be generalizable to other computational cognitive architectures, at least with respect to the key points about likelihood.

This primer is focused on the procedural and mathematical steps to implement likelihood-based model validation and selection. However, recognizing that code examples might be useful, we have made an accompanying Jupyter notebook available at **https://github.com/UWCCDL/MLExACTR/**. The notebook provides step-by-step instructions as well as runnable Python examples containing all of the data and code to generate the results and figures in this paper.

## Theory

### Overview of ACT-R

The ACT-R architecture is a substantive theoretical proposal regarding the neurofunctional structures and processes that enable the human mind [35]. It is implemented in freely distributed, open access software that defines the modules, buffers, and parameters available to the architecture. Although the details of the implemented theory continue to evolve, the architecture itself is functionally static. It is not capable of replicating or predicting human performance until knowledge and skill are added to produce one or more models.

ACT-R models are controlled through the interaction of two types of memory, declarative and procedural, as well as their associated parameters. Declarative memory represents the model's *knowledge*; it is represented as record-like structures called *chunks* and is used to encode semantic and episodic memories, perceptual inputs, and motor commands. Chunks are maintained and processed in specialized modules, whose computational activity has been associated with distinct brain regions [36,37].

Procedural memory, on the other hand, encodes the model's *skills*. It is represented by condition–action pairs known as production rules or simply *productions*, which encode primitive visuomotor operations (such as "IF you see a light, THEN press the button") as well as the specific task operations ("IF you have decided to examine a visual feature, THEN collect in working memory the values of the feature across the first row"). Productions are managed by the procedural module which abstracts the function of the human striatum [36,38,39].

Cognition arises from the interaction of these two forms of memory, enabled and constrained by the structure of the architecture, with production rules either executing habitual cognitive and motor actions for reflexive responses or retrieving memories for more deliberate and elaborate behaviors. Chunks and productions interact through module-specific buffers: each module produces or elaborates information in the form of chunks and makes them available to production rules by placing them into buffers. Productions control behavior by transferring information (i.e., whole or partial chunks) between buffers, where they can be processed appropriately. A special retrieval module is responsible for retrieving chunks from long term semantic and episodic memory.

Both productions and chunks are selected serially: At any given time, only one chunk can be retrieved from declarative memory and only one production can be executed by the procedural module. Production rules are selected based on their *utility*, a scalar quantity associated with each production that summarizes the previous history of rewards that have resulted from selecting that production. When a reward is given, the utility of a production rule that has been fired is updated according to the reinforcement-learning-like rule:

$$U_n(p) = U_{n-1}(p) + \alpha[U_{n-1}(p) - (R - \Delta t)]$$ \hfill (3)

Where $r_t$ is the reward given at time $t$ and $\Delta t$ is the time that has elapsed from the $n$-th firing of $p$ to time $t$. The selection of production rules is made non-deterministic by adding noise to its utility, whose standard deviation is controlled by a specific parameter $\tau$.

Episodic and semantic chunks also have an associated scalar quantity, called *activation*. If a production's utility reflects the history of rewards, a chunk's activation reflects the history of encoding and retrieval of that information. This history is condensed as the sum of power decaying *traces* of previous encounters with that knowledge. Specifically, the activation $A_t(c)$ of chunk $c$ at time $t$ is given by:

$$A_t(c) = \log \sum_i (t - t_i)^{-d}$$ \hfill (4)

As in the case of productions, the selection of a chunk is also made non-deterministic by addition of noise to its activation. The standard deviation of the noise is controlled by the *s* parameter.

The interaction of actions and memory retrievals leads ACT-R models to produce behavioral responses. These responses also have associated *response times*. Specifically, all procedural actions, in ACT-R, take a fixed amount of time, such that the time to execute a production $p$ is fixed and equal to a constant action time $t_A$, which by default is set to 50ms.

Memory retrievals, on the other hand, take a variable amount of time that depends on the activation of the retrieved chunk. The time $T(c)$ needed to retrieve a chunk $c$ is:

$$T(c) = F e^{-A(c)} \tag{5}$$

where $F$ is a *latency factor* that scales the degree to which the activation of a chunk affects the speed of its retrieval.

Although ACT-R contains dozens of additional parameters, the most widely used ones are those that regulate the function of its declarative and procedural memory, as described in Equations 3-5. Table 1 provides an overview of their names, their computational meaning, the type of memory they regulate, and the behavioral measure they affect.

**Table 1**: Overview of the main ACT-R parameters.

| Memory | Affected Measure | | Parameter | ACT-R Name | Meaning (Equation) |
| --- | --- | --- | --- | --- | --- |
| | Accuracy | RT | | | |
| Declarative | • | • | $d$ | :BLL | Memory trace decay (Eq. 4) |
| | • | • | $s$ | :ANS | Activation noise |
| | | • | $F$ | :LF | Latency factor (Eq. 5) |
| Procedural | • | | α | :ALPHA | Learning rate (Eq 3) |

| | | | | |
|---|---|---|---|---|
| | • | | $\tau$ | :EGS | Procedural noise |
| | | • | $t_A$ | :AT | Action time |

**Understanding Likelihood**

Having introduced ACT-R, we can now dig deeper into likelihood. As noted above, the likelihood of a particular model $m$ with a vector of parameters $\boldsymbol{\theta}$, $\mathscr{L}(m, \boldsymbol{\theta} \mid \mathbf{x}_o)$ can be interpreted as the plausibility that the model would produce that observed data $\mathbf{x}_o$:

$$\mathscr{L}(m, \theta \mid \mathbf{x}_o) = P(\mathbf{x} = \mathbf{x}_o \mid m, \boldsymbol{\theta}) \tag{1}$$

There are three components that play a role in the likelihood calculation. The first is the model $m$ and its free parameters $\boldsymbol{\theta}$. Strictly speaking; likelihood applies to *parametric* models, that is, probability distributions characterized by equations; the parameters of the models are the parameters of their equations. Models developed within a cognitive architecture typically contain a mixture of parametric and non-parametric components. In ACT-R, for example, chunks and productions are discrete, symbolic, non-parametric components, while the equations that govern their selection, decay, and retrieval are parametric. Although both parametric and non-parametric aspects of cognitive architectures affect the behavior of the model, likelihood applies only to the specific parameters that govern the parametric components (e.q., Table 1).

The second component is the observed data, $\mathbf{x}_o$. The observed data are given as fixed data points that the model is trying to generate. They represent the yardstick against which the model's performance is measured, and their variability is not considered.

The second is the *distribution* of a model's predictions. The predictions are the model's output data, that is, the simulated behavioral responses and response times. Unless a model is completely deterministic, for any set of parameters $\boldsymbol{\theta}$ there will be some variability associated with the model's predictions. In ACT-R, this variability comes from two parameters, one that controls the noise in chunk activation levels ($s$) and one that controls the noise in the utility value of productions ($\tau$). The variability generated by these parameters is desirable, as it is key to making the calculation of Equation 1 possible. The variability of a model's predictions can be expressed in the form of a probability density function $P(\mathbf{x} \mid m, \boldsymbol{\theta})$. In most cases, this probability density function can be approximated as a normal distribution with a mean and a standard deviation. For more complicated cases, a probability density function can be estimated empirically through a distribution-fitting procedure, such as kernel density estimation [40]. Packages to perform kernel density estimation are available for both the R and the Python programming languages.

The likelihood of a model can thus be interpreted as the probability that the model would output data that is identical to the observed value: $P(\mathbf{x} = \mathbf{x}_o \mid m, \boldsymbol{\theta})$. The relationship between these quantities is shown in Figure 1, in the simple case in which the data and the model predictions pertain to a single variable, $x$.

For the rest of this paper, we will assume that "data" refers only to the observed data, and we will use the shorthand $P(\mathbf{x} \mid m, \boldsymbol{\theta})$ to refer to $P(\mathbf{x} = \mathbf{x}_o \mid m, \boldsymbol{\theta})$. This will make most of our notations less cumbersome. Please bear in mind, however, the probability densities used to calculate likelihood refer to the space of the *possible data* generated by the model, against which the observed data **x** is matched. The space of possible data is, in itself, an important element to consider when evaluating the plausibility of a model [41,42].

In most cases, the observed data $\mathbf{x}_o$ is composed of multiple variables $x_1, x_2, \ldots x_n$; in these cases, the term $P(\mathbf{x} \mid m, \boldsymbol{\theta})$ represents the value of their *joint* probability distribution. It is important to stress this because, in most cases, the observed variables are not independent of each other, and their relative correlations affect the estimation of the joint probability distribution. For instance, response times and accuracies are often negatively correlated due to the speed-accuracy tradeoff; and different measurements on a learning curve are correlated because each measurement usually depends on the previous. Figure 3 provides an illustration of how the joint probability distribution of two variables $x$ and $y$ with identical means and standard deviations change as a function of their correlation, from $r = -0.5$ to $r = 0.5$. It is easy to see that, although both variables have the same probability of assuming the value of 2, the joint probability of $x = y = 2$ changes as a function of their correlation.

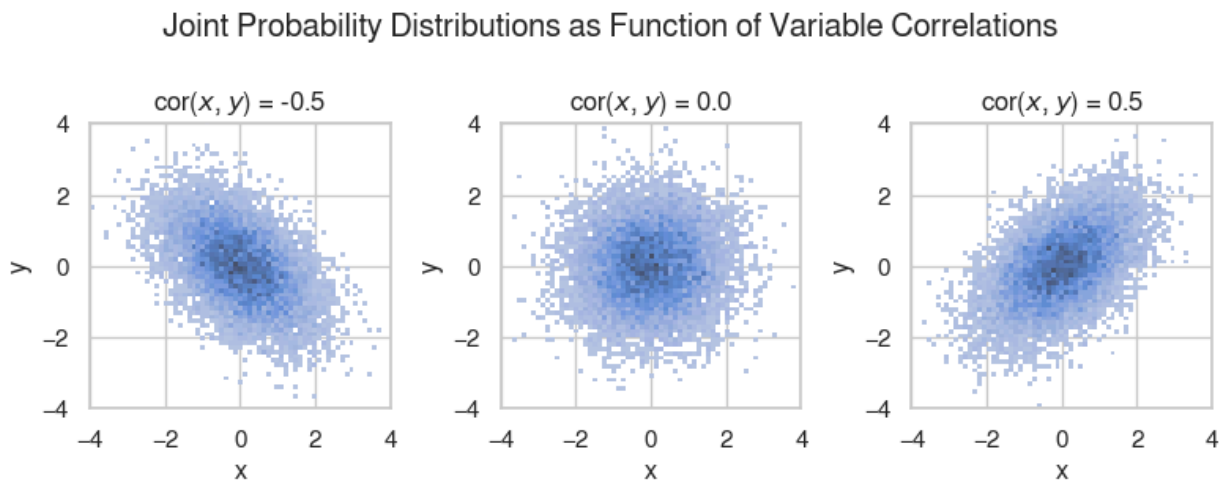

**Figure 3**: Joint probability distributions of two variables x and y, both of which are normally distributed with a mean $M = 0$ and a $SD = 1$, but with different correlations.

When variables are independent (e.g., their correlation is zero) or we can safely assume that they are, Equation 1 can be further simplified: the joint probability becomes the product of their independent probabilities, and the equation can be rewritten as:

$$\mathscr{L}(m, \theta \mid \mathbf{x}) = \prod_i P(x_i \mid m, \boldsymbol{\theta}) \tag{6}$$

In practice, some of these probability values will be exceedingly small, especially for bad models or difficult-to-fit data. For this reason, it is common to measure a model's likelihood $\mathscr{L}$ in terms of *log*-likelihood instead:

$$\log \mathscr{L}(m, \theta \mid \mathbf{x}) = \log \prod_i P(x_i \mid m, \boldsymbol{\theta}) = \sum_i \log P(x_i \mid m, \boldsymbol{\theta}) \tag{7}$$

In Equation 7, vanishingly small products of probabilities become sums of negative numbers, significantly reducing the rounding errors.

**Probabilities, Likelihoods, Generative Processes, and Inference**

Equation 1 captures the relationship between the likelihood of a model and its parameters and the probability of the data it generates. Conceptually, the difference between likelihood and probability can be understood as the difference in the direction of the connection between models (and parameters) and data. Probabilities link the specific model and its parameters to the observable *data*: the model hypothesizes a generative process that, for a given set of parameter values, defines a probability distribution over the data $\mathbf{x}$. Within this framework, a probability can be assigned to observing the specific data point **x₀** according to this distribution.. Likelihoods, however, move in the opposite direction, and connect the observed data to the parameters of the model, thus permitting legitimate *inferences* from the data about the *most likely* model parameters.

Figure 4 illustrates this distinction in reference to a simple model that is familiar to the reader: the normal distribution function. Given specific values of its mean μ and standard deviation σ, the function will assign a probability to the observed data **x**$_o$. And given some observed values **x**$_o$, likelihood can be used to select the most likely values of μ and σ.

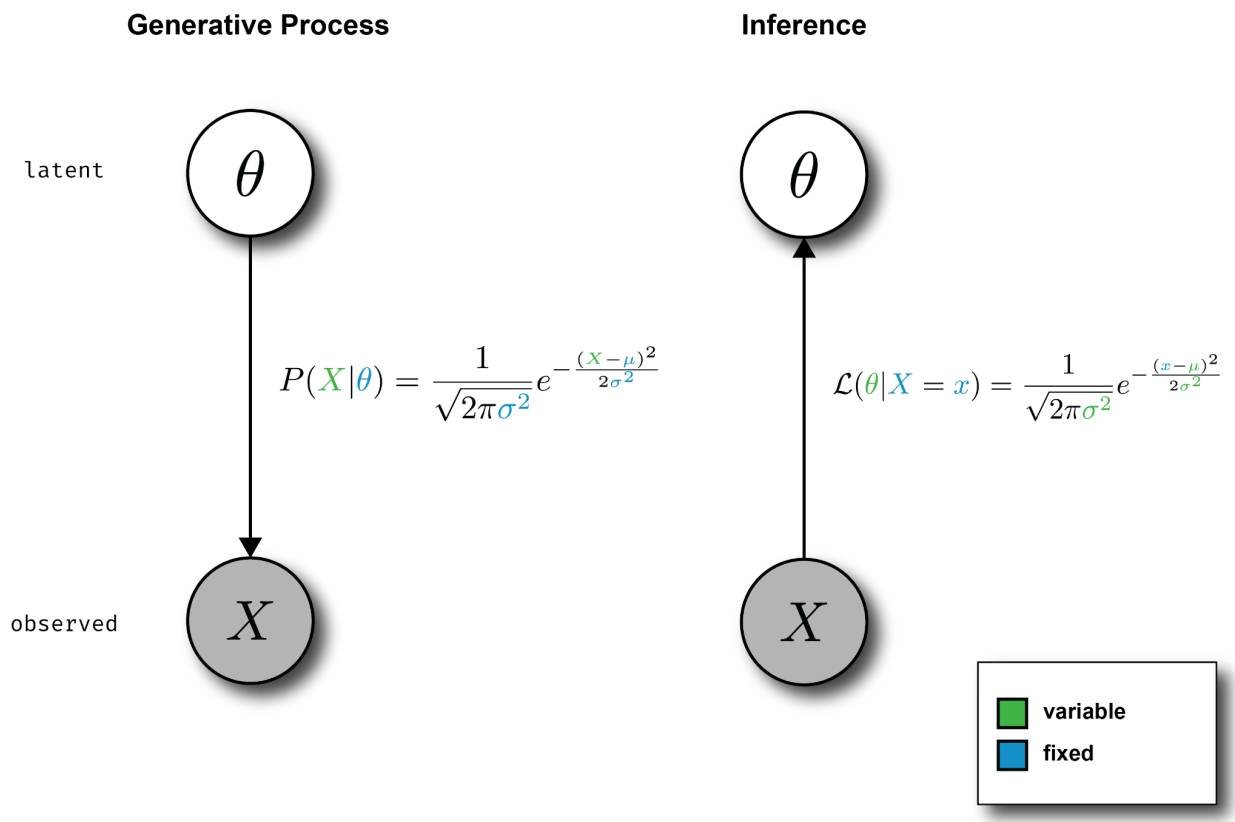

**Figure 4**: Probabilities and likelihoods connect model parameters and data in the opposite direction.

**Using Likelihood to Fit Models: Maximum Likelihood Estimation (MLE)**

To provide concrete examples for the concepts discussed in this paper, we will consider ACT-R models for a simple task: a decision-making paradigm known as the Incentive Processing Task or IPT [43]. The IPT belongs to the class of paradigms known as two-alternative, forced-choice (2AFC) tasks, which are commonly used in psychology experiments. Many modeling frameworks, such as drift-diffusion modeling [44], are explicitly designed for 2AFC tasks. The IPT is also one of the tasks used in the Young Adult version of the Human Connectome Project [45], a large repository of neuroimaging and behavioral data from over 1,200 individuals who have performed this task (and many others) while in a fMRI scanner. The example data shown here comes from two participants from the HCP dataset.

In the IPT, participants are asked to guess if the number behind a mystery card (represented by a "?", and ranging from 1 to 9) is larger or smaller than 5. Participants respond by pressing one of two buttons mapped to the "More" or "Less" option; they win money when making correct guesses and lose money for incorrect ones. After making a guess, participants are shown the real value of the number, together with an arrow pointing in the direction of the correct answer (up for "More", down for "Less", and a double-headed arrow when the number is exactly 5). The feedback could be of three types: "Reward" (the arrow is green, participants win $1), "Punishment" (the arrow is red, participants lose –$0.50), and "Neutral" (the double-headed arrow is gray, and participants do not win nor lose any amount). Critically, the feedback does not depend on the subject's response, but is determined in advance; the sequence of pre-defined feedback is *identical* for all participants. In the data used here from the HCP dataset, the task was presented in two runs, each of which contains 32 trials divided into four blocks. Blocks could be "Mostly Lose" (6 loss trials pseudo-randomly interleaved with either 1 neutral and 1 reward trial, 2 neutral trials, or 2 reward trials) or "Mostly Win" (6 win trials pseudo-randomly interleaved with either 1 neutral and 1 loss trial, 2 neutral trials, or 2 loss trials). In each of the two runs, there were two "Mostly Win" and two "Mostly Lose" blocks. All participants received money as compensation for completing the task, and the amount of reward is standard across subjects.

Figure 5 depicts the raw series of decisions and the corresponding feedback for two participants. For each participant and each run, the graph depicts the timeline of "more" or "less" guesses as they occurred. Each choice is color-coded according to the feedback they received. Note that, because in this task the feedback schedule was predetermined, the colors of all choices in the same position are the same.

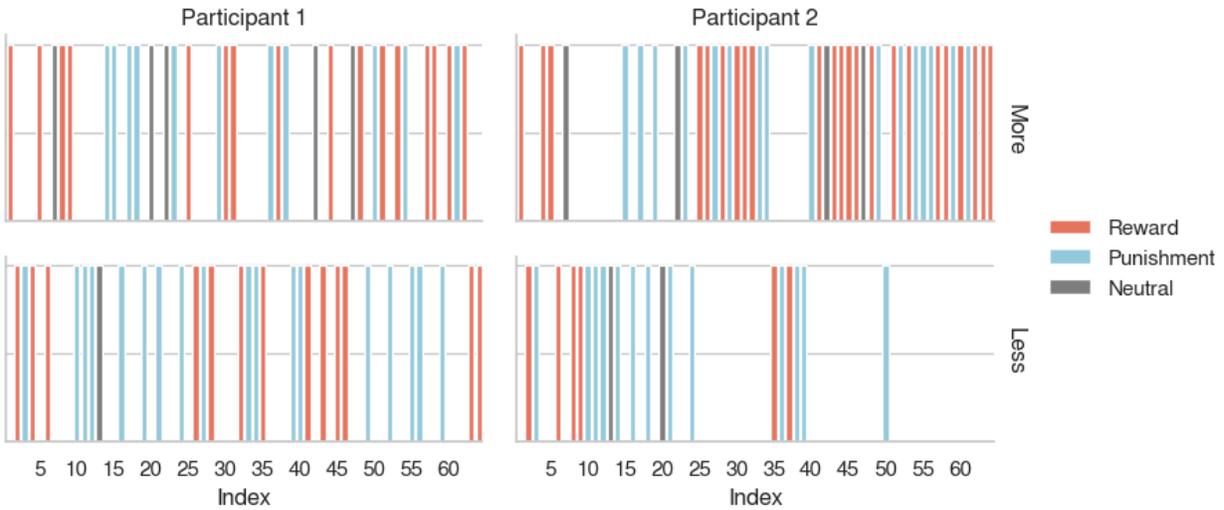

**Figure 5**: Raw series of decisions and corresponding feedback for two participants.

Because the feedback schedule is predetermined and independent of choices, participants always have the exact same number of wins and losses. As a consequence, raw accuracy is not a meaningful measure of performance. Following Yang, Sibert, and Stocco [46], we will analyze the data in terms of "Stay Probability", that is, the probability of repeating the same choice after feedback. Figure 6 provides the stay probabilities for each type of feedback (reward, punishment, and neutral) as well as the global stay probability for each participant.

In the following sections, we will introduce different ACT-R models for this task, and examine how likelihood can be used to evaluate the relative goodness-of-fit of those models to the data from these two individuals.

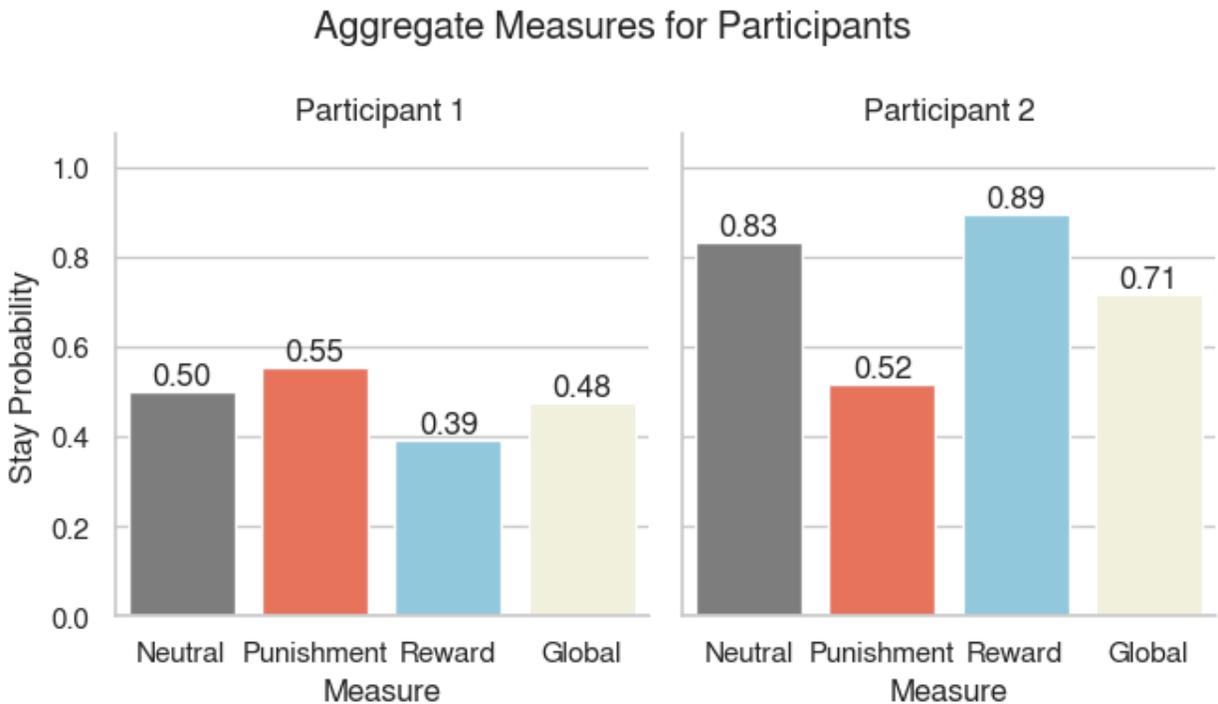

**Figure 6**: Aggregated behavioral measures (stay probabilities) for the two participants in the Incentive Processing Task.

**Deriving Likelihood from Empirical Probabilities Through Monte Carlo Simulations**

As a first step, we will focus on the global stay probability (beige bars in Figure 6) and will consider a simple ACT-R model that relies on memory of previous choices to make a decision. After making any choice in the IPT, the model memorizes what action was taken (i.e., "More" or "Less") and its associated result ("Reward", "Punishment", or "Neutral"). When making the next decision, the model retrieves the most active memory of previous choices. If the memory was of a choice previously associated with a "Reward" feedback, the model repeats the same choice. If not, the model performs the opposite choice. This memory-based approach to decision-making is common in the ACT-R literature, and is known as Instance-Based Learning [47,48]. Since the model relies only on declarative memory, it was implemented in PyACTUp, a standalone version of ACT-R's declarative memory written in Python.

The simplest way to estimate the probability that a model would reproduce the behavior of Participants 1 and 2 is to run the model multiple times with a given set of parameters θ and record the distribution of results. Figure 7 shows the results of simulating 5,000 runs of the model with parameters $d = 0.5$ and $s = 0.25$. The histogram represents the observed empirical distribution of stay probabilities, while the black line represents its probability density (estimated with a kernel-density estimator). The orange and purple points represent the position of participants 1 and 2 within this distribution; their coordinates on the $x$-axis represent their observed stay probabilities and their coordinates on the $y$-axis represent their corresponding likelihood values $P(\mathbf{x} \mid m, \boldsymbol{\theta})$.

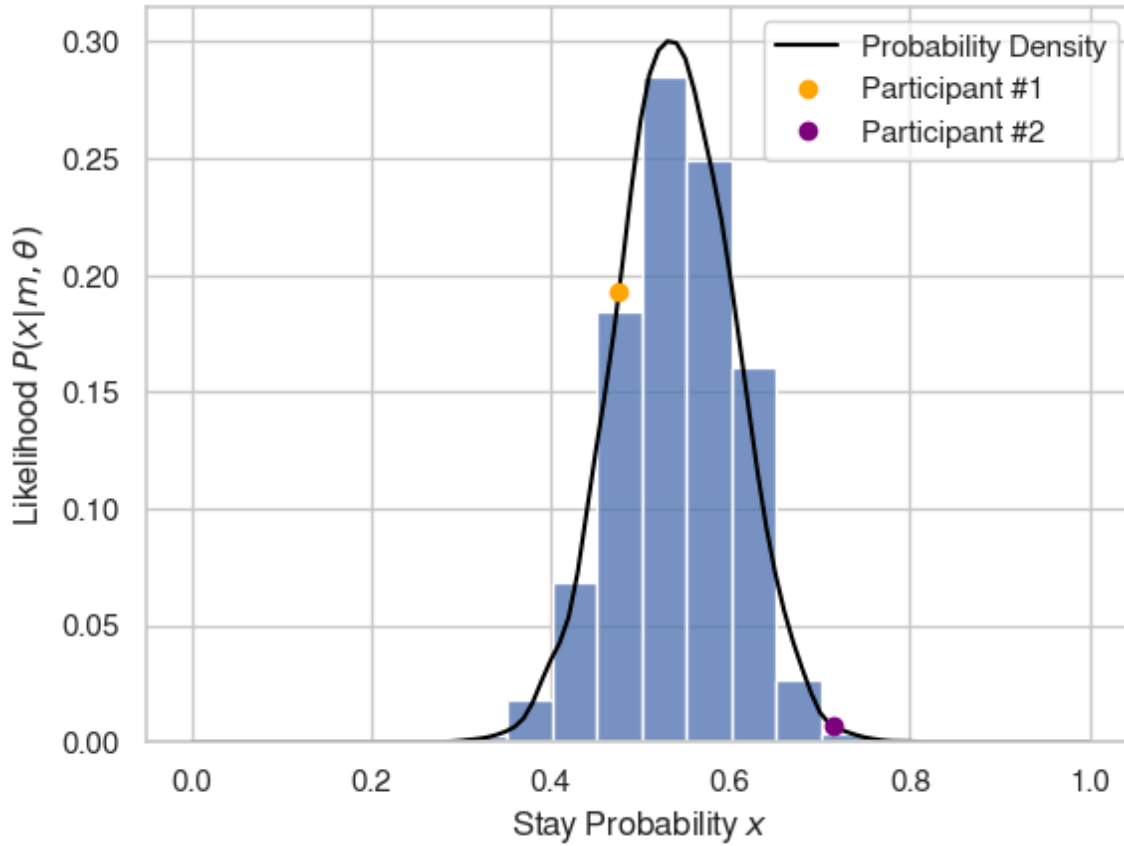

**Figure 7**: Distributions of observed stay probabilities obtained from running the model 5,000 times (blue bars), together with the KDE-estimated probability density (black line). The orange and purple dots represent the stay probabilities of Participants 1 and 2 (*x*-axis) and the likelihoods that the model would reproduce their data (*y*-axis). See Figure 3 for comparison.

**Exploring the Parameter Space**

Figure 7 suggests the model, with its default parameters $d = 0.5$ and $s = 0.25$, already does a decent job at fitting the behavior of participant 1. But is there a set of parameters that *maximizes* the probability of reproducing their data? Stated in the other direction, given the data we have for Participant 1, what are the most likely architectural parameters for that person?

To answer this question, one must explore the possible combinations of $d$ and $s$, that is, the *parameter space* of the model. A simple way to do so is to use a "grid search" approach: systematically varying the values of $d$ and $s$ in fixed intervals and recording the likelihood values for the data of participant 1 for each combination. Figure 8 depicts the results of this full grid search approach for different combinations of $d$ and $s$: the values of both parameters are parametrically varied between 0 and 1 in increments of 0.05 and the recorded likelihood values of $P(\mathbf{x}_1 | m, \boldsymbol{\theta})$ are represented as different cell colors.

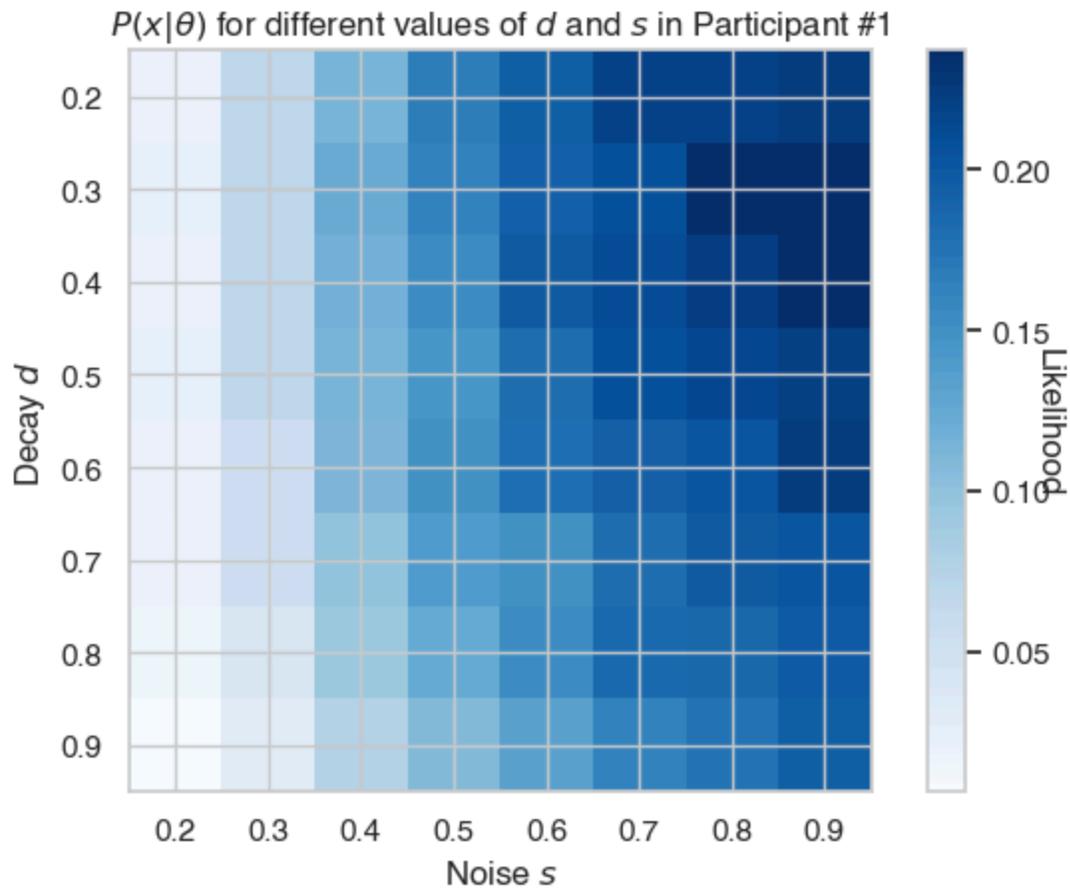

**Figure 8**: A 2D representation of how the likelihood of reproducing the data of Participant 1 varies with different values of *d* and *s*.

Running simulations in full grid search can be computationally prohibitive due to the combinatorial explosion occurring as the number of parameters and resolution of the search increase. For this reason, it is usually  advisable to use optimization techniques. Because we do not have analytic expressions for how the parameters translate into stay probabilities, we must use a derivative-free method, such as Powell's method [49] or the Nelder-Mead simplex method [50], which are more efficient in exploring the parameter space and are included in most data science software packages.

In this tutorial, we will use Powell's method as an example. Although more precise or more efficient algorithms have been proposed [51,52], Powell's method has the advantage of being a textbook, off-the-shelf algorithm that is included in most common software packages. In its simplest form, the method searches the space along each parameter separately. When only two parameters are considered, the space is bidimensional and Powell's method often results in a series of vertical and horizontal adjustments. The path explored by Powell's method in the data is shown in Figure 9 as a series of points of increasingly greater opacity from the first point sampled to the last. The final result is shown as a red star: According to this method, the best-fitting parameters are $d = 0.47$, $s = 0.65$.

A visual inspection of Figure 9 suggests that, in this case, Powell's seems to have landed on a combination of parameters that yields a decent fit but does not seem to be optimal. This is a side effect of combining derivative-free methods with likelihood estimates derived from multiple simulations: empirically-derived estimates tend to be noisy, which means that the parameter space is not smooth, which might lead optimization techniques to end in a local maximum.

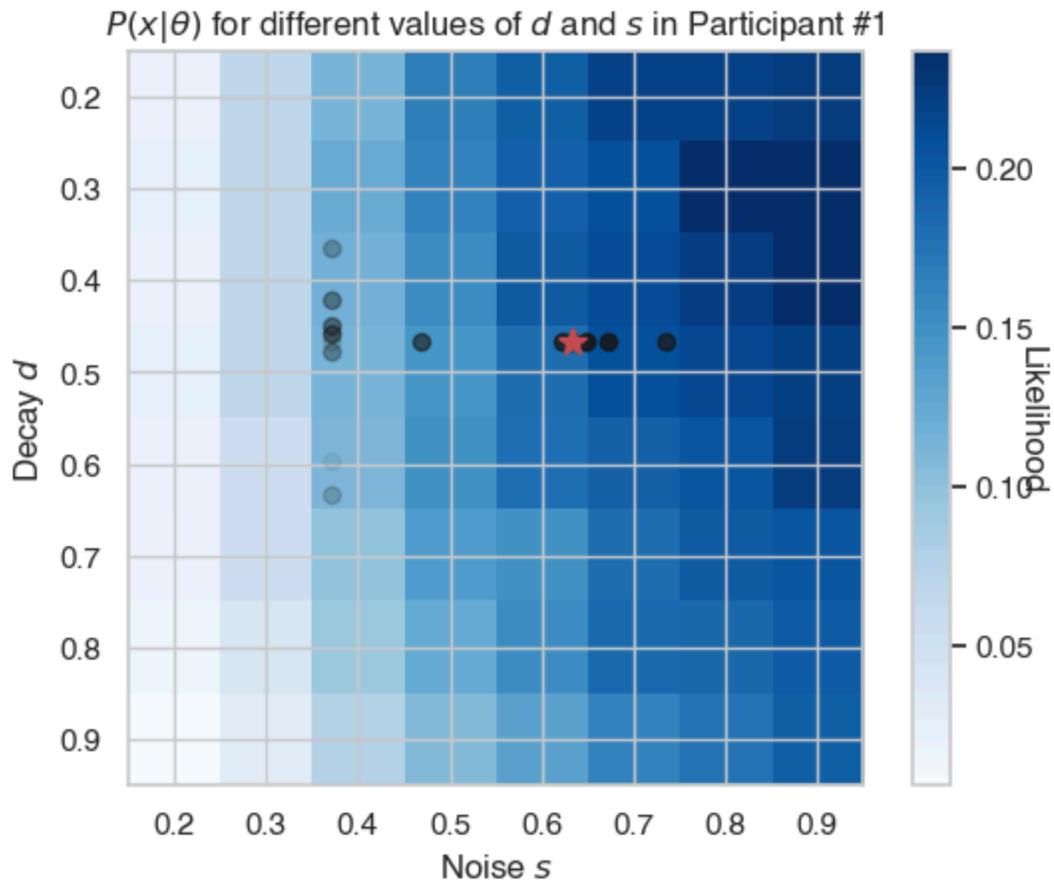

**Figure 9**. An example of parameter space exploration with Powell's method. Dark points represent different locations in parameter space explored by the algorithm; the red star represents the combination on which the algorithm converged. The statistical nature of the estimates results in a parameter space that is not smooth and might lead the algorithm to suboptimal results.

**Fitting a Model Using Multiple Measures**

Cognitive researchers rarely focus on a *single* aggregate behavioral measure; more often than not, any task yields two or more measures of interest. For example, an experimenter developing a model for the Stroop task would be interested in response times for congruent and incongruent trials. Similarly, researchers using the IPT might be interested in examining an individual's stay probabilities following successful ("reward") and unsuccessful ("punishment") trials. If an individual is following a simple "Win stay, lose shift" strategy, for example, then they would exhibit higher stay probabilities after reward than after punishment trials.

The principles described above work for any number of variables. However, remember that the definition of likelihood is based on the *joint* probability distribution of the data. When running simulations, the joint probability distributions can be estimated directly from the data. The left panel of Figure 10 plots the results of a portion of the parameter space of Figures 8 and 9, showing the results of 5,000 simulated runs of the model with the optimized parameter values of $d = 0.47$ and $s = 0.65$ across two axes corresponding to the stay probabilities after a reward and after a punishment. The orange dot represents the performance of Participant 1.

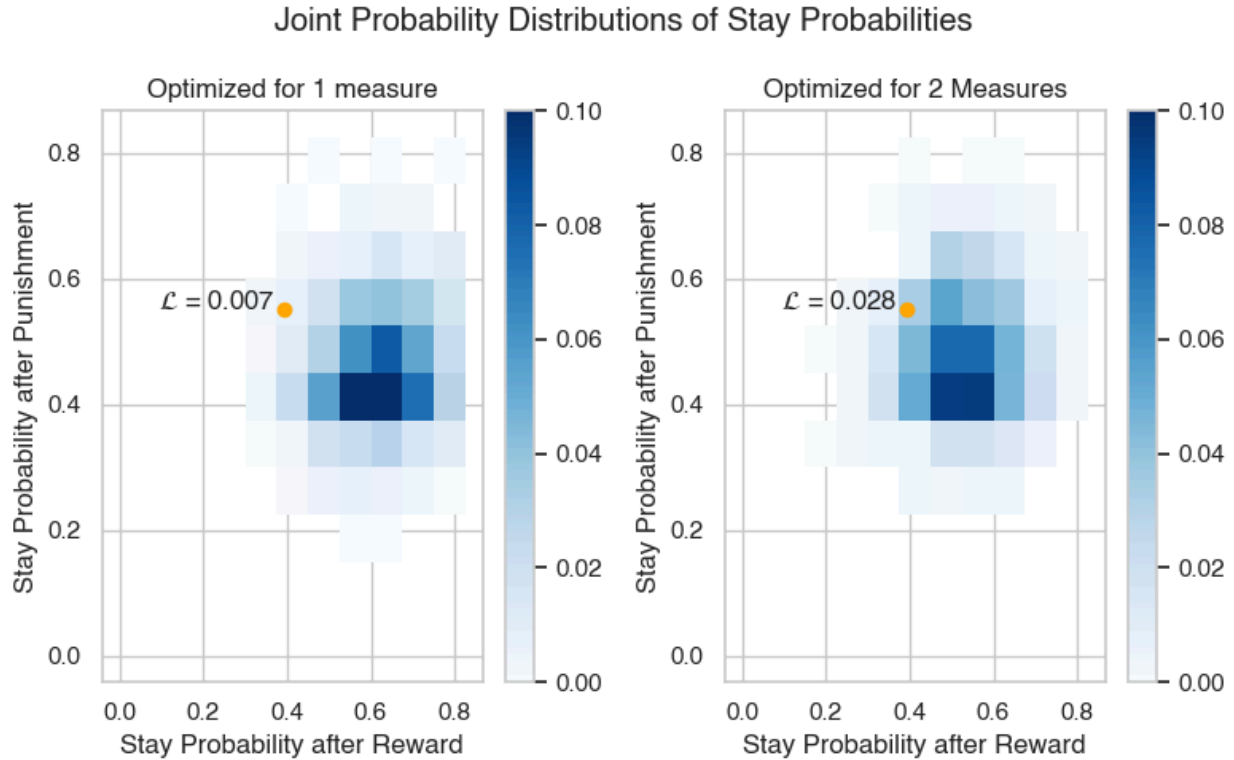

**Figure 10**: Joint probability distribution that the model would generate specific combinations of stay probabilities after reward and punishment trials when optimized for one measure (global stay probabilities: left) or for two measures (stay probabilities after reward and after punishment: right). The orange dots represent the combinations of stay probabilities for Participant 1.

As the plot shows, once both measures are considered, the model's fit to Participant 1's data is still poor even *after* the optimization process. The figure also makes it clear *why*: the poor fit is due to the peculiar nature of Participant 1's performance, where they are more likely to stay after a "punishment" than after a "reward" feedback. So, while it is possible to get the model to reproduce the *average* stay probability across all trials (an aggregate measure that includes the neutral trials), the peculiar distribution of stay probabilities after rewards and after punishments is harder to replicate.

It is possible to use the same optimization approach (and, in our online notebook, virtually the same code) to fit the model to Participant 1 across *both* measures: in this case, we computed the joint probability distribution of the two stay probabilities (using a kernel density estimator) for each combination of parameter values in the model. The right-hand side of Figure 10 shows the results of this second optimization attempt, based on two measures instead of one. In this case, the optimization process has converged on a lower decay value ($d = 0.12$) and a higher noise value ($s = 0.75$). As Figure 10 shows, these new parameters produce a different distribution of stay probabilities, one in which the model has a greater likelihood ($\mathscr{L} = 0.028$ instead of $\mathscr{L} = 0.007$) to produce the specific combination of stay probabilities of Participant 1. Note, however, that likelihood has only increased slightly, reflecting the fact this model simply is not a good account of a participant whose choice behavior is biased toward "win shift, lose stay."

**Assuming Independent Measures**

Sometimes, the joint probability distribution of a variable might not be available. As a necessary approximation, one could assume that the distributions of the two variables are uncorrelated; this simplifying assumption is akin to the assumption of independence in the Naive Bayes classifier. In this case, the joint probability is just the product of the probabilities, and the joint log-likelihood is the sum of the log likelihoods of the individual variables.

If the correlation between the two measures $x$ and $y$ is known or can be calculated, then we can calculate a covariance matrix $\mathbf{\Sigma}$:

$$\Sigma = \ [\sigma_x \times \sigma_y \times r(x,y), \quad \sigma_x^{\,2}$$

$$\sigma_y^{\,2}, \qquad \sigma_x \times \sigma_y \times r(x,y)]$$

Once we have the covariance matrix, we can use it to calculate the multivariate normal distribution for $x$ and $y$: $P(x, y) \sim N(\mathbf{u}, \mathbf{\Sigma})$. The joint probability distribution now successfully models the observed pattern of results.

**Using Likelihood on a Trial-by-Trial Basis**

So far, this paper has only considered *aggregate* measures of behavior. Aggregate measures, however, suffer from considerable pitfalls, in that they risk concealing important information about inter- and intra-individual differences [53]. They are also sensitive to outlier values. Consider the time course of response times depicted in Figure 11[1], which come from the responses of a participant to different memory cues. It is clear that the third trial is an outlier, taking almost 20x as long as any other trial in that session. Such outlier trials are common in real experimental data. The problem is, of course, that that particular outlier observation has a huge impact on the global RTs: in this case, the mean RT including the outlier (18.4s) is almost twice as much as the mean RT of all of the other trials (10.9s).

---

[1] Data from Hake et al. (2023).

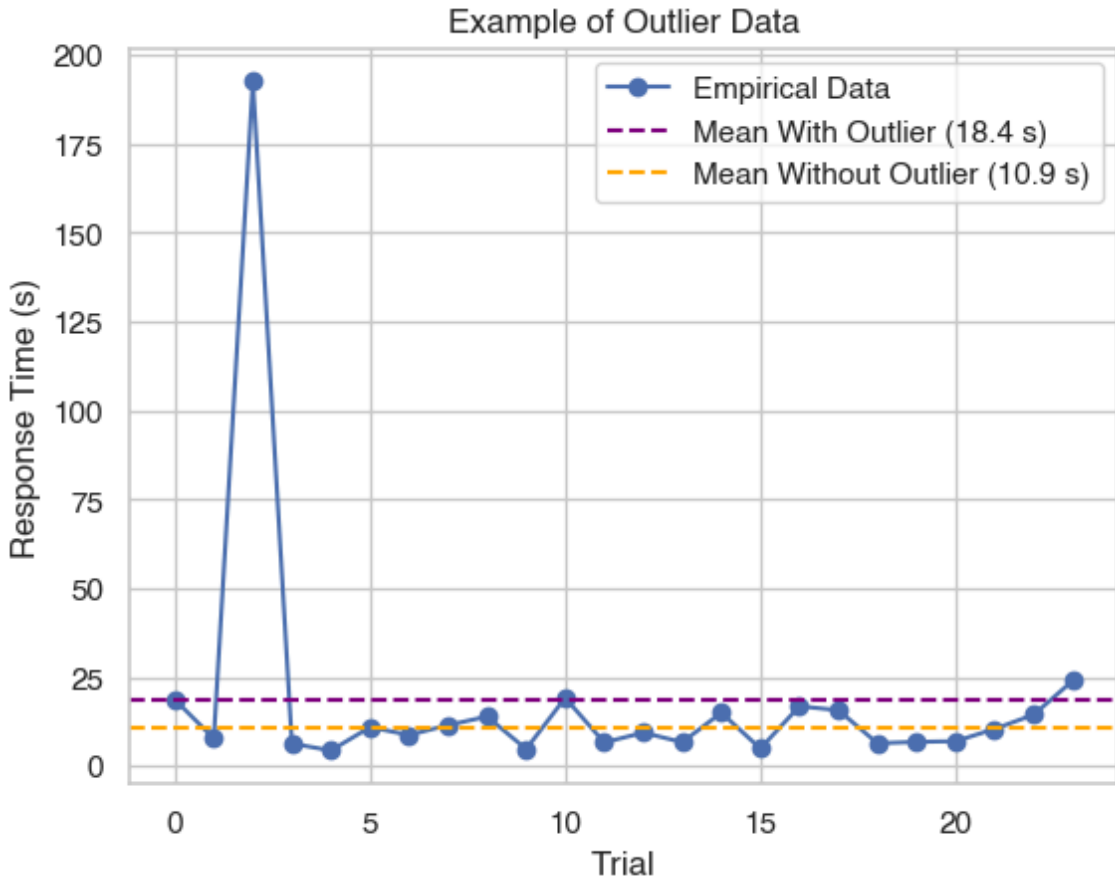

**Figure 11**: A series of response times from one participant from Hake et al's (2023) experiment, including an apparent outlier response in the third trial.

Of course, it is possible to remove outlier observations, but that requires a series of arbitrary assumptions—what is, exactly, the threshold for an outlier? And how about abnormally *fast* response times? Instead of defining arbitrary thresholds, MLE handles outlier response times naturally and nicely: any unusually slow or unusually fast response time will simply be proportionally *less likely*. At the same time, trial-by-trial analysis allows a researcher to take full advantage of all of the possible observations in the data, gathering a full picture of an individual's behavior. For these reasons, trial-by-trial estimation is often used and recommended in other modeling paradigms [11].

Depending on the circumstance, however, there are some disadvantages. For example, in trial-by-trial analysis, one can typically observe only the specific response time and choice made in that particular instance. Experimental studies often focus on *derived* measures, such as the differences between different groups of trials (congruent vs, incongruent response times in the Stroop task). For example, the mean probability of switching after a reward or punishment, which was used in the previous example, is not a measure that exists when looking at the data at the granularity level of trial-by-trial analysis.[2]

Another disadvantage is that trial-by-trial analysis requires explicit, closed-form expressions for the probability distributions of responses and response times. As the next sections will show, these probabilities are easy to derive for simple models. But, as the models become more complex, these expressions are much more difficult to derive and depend on the specifics of the model. In these cases, relying on simulations of aggregated data might be the simplest and safest option.

---

[2] Of course, one could work around that. For example, one could fit the data by computing the trial-by-trial probability of switching. Or one could have a model that behaves differently after rewards and punishments.

If the model is relatively simple, however, as it is the case of *most* ACT-R models for laboratory tasks, then one can take full advantage of trial-by-trial analysis.

**Trial-by-Trial Analysis in ACT-R**

In trial-by-trial data, the vector of data $\mathbf{x}$ is a vector of points that were acquired at consecutive times $t = 1, 2, \ldots, N$.

$$\mathscr{L}(m, \theta \mid \mathbf{x}) = P(\mathbf{x} \mid m, \boldsymbol{\theta}); \qquad \mathbf{x} = \{x_{t=1}, x_{t=2}, \ldots x_{t=N}\}$$

In this case, the joint probability distribution of the data is best understood as a conditional distribution of observing the data point $x_t$ at times $t$, given that we have observed *all* of the previous data points at times $t-1, t-2, \ldots, 0$. Following this logic, our probability of observing the full series of data $\mathbf{x}$ can be expanded as the probabilities of observing each data point, conditional on all of the preceding ones:

$$P(\mathbf{x} \mid m, \boldsymbol{\theta}) = P(x_1 \mid m, \boldsymbol{\theta}) \cdot P(x_2 \mid m, \boldsymbol{\theta}, x_1) \cdot \ldots \cdot P(x_N \mid m, \boldsymbol{\theta}, x_1, x_2, \ldots x_{N\text{-}1}) \qquad (8)$$

This leads to a series of conditional probabilities. Fortunately, ACT-R's behavior can be approximated as a *single-step* Markov model, and its future behavior is determined solely by its current state (as embedded in the current state of its knowledge, buffers, and parameters) and not by its previous history of choices. The previous history of interaction with the environment can have an effect on the model only to the extent by which it is captured by the model's knowledge or parameters. For example, a model could make decisions based on memories of previous actions, or be more likely to retrieve certain memories after having previously seen them, or more likely to select an action after experiencing a reward. So, if the model's knowledge is correctly updated after every choice, the temporal dependencies between choices can be ignored, and Equation 1 becomes:

$$\mathscr{L}(m, \theta \mid \mathbf{x}) = P(\mathbf{x} \mid m, \boldsymbol{\theta})$$

$$= P(x_1 \mid m, \boldsymbol{\theta}) \cdot P(x_2 \mid m, \boldsymbol{\theta}) \cdot \ldots \cdot P(x_N \mid m, \boldsymbol{\theta})$$

$$= \prod_i P(x_i \mid m, \boldsymbol{\theta}) \tag{9}$$

And the global log-likelihood becomes the sum of each trial's log-likelihood:

$$\log \mathscr{L}(m, \theta \mid \mathbf{x}) = \log \prod_i P(x_i \mid m, \boldsymbol{\theta})$$

$$= \sum_i \log P(x_i \mid m, \boldsymbol{\theta}) \tag{10}$$

This trial-by-trial approach is analogous to the already-established *model tracing* technique in used in cognitive tutors (Koedinger & Anderson, 1993), whereby an ACT-R model is artificially forced to make the same decisions as the student it is trying to model before making a prediction for the next choice.

### Analytical Definitions for Accuracies and Response Times

It is highly impractical to collect trial-by-trial probability densities using Monte Carlo simulations. So, it is *de facto* necessary to use analytic definitions that specify the probabilities that a specific procedural action or memory chunk is selected and the probability densities of their associated response times. Fortunately, they are not too difficult to calculate.

In the case of a procedural system, the probability of selecting a production $p$ is a softmax function of its utility:

$$P(p) = e^{-U(p)/\tau} / \sum_i e^{-U(i)/\tau} \qquad (11)$$

where $\tau$ is a parameter that controls the noise associated with the selection (:EGS in Table 1).

Similarly, In the case of the declarative system, the probability $P(c)$ of retrieving a specific chunk $c$ is a softmax function of its activation:

$$P(c) = e^{-[A(c)-\tau]/s} / \sum_i e^{-[A(i)-\tau]/s} \qquad (12)$$

where s is a parameter that controls selection noise in declarative memory (:ANS in Table 1).

The probability distribution function for accuracy is defined over a discrete space–accuracies are either 1 or 0. The probability of a correct or incorrect answer is given by $P(c)$ and $1 - P(c)$ respectively.

The response time associated with retrieving a chunk is a deterministic function of that chunk's activation at the time of retrieval (Equation 5). Specifically, behavioral responses that depend on memory retrieval will have a response time $RT$ of the form:

$$RT = t_0 + Fe^{-A(m)} \tag{13}$$

where $t_0$ is a term that absorbs perception/motor times. The source of variability in $RT$ is given by the activation noise $s$. Because noise is distributed as a logistic function with standard deviation $\pi s/\sqrt{3}$, the activation $A(c) + s$ is also distributed as a logistic distribution with mean $\mu = A(c)$ and $\sigma = \pi s/\sqrt{3}$. Therefore, $t_0 + Fe^{-A(m) + s}$ is a shifted log-logistic distribution (also known as a Fisk distribution) with parameters $\alpha = e^{\mu} = e^{-A(m)}$, $\beta = 1/\sigma = \sqrt{3}/\pi s$:

$$P(t) = [\ \beta\ /\ (F\alpha\ )]\ [(t - t_0)\ /\ \alpha]^{\beta-1}\ /\ (1 + (t - t_0)\ /\ F\alpha)^{\beta})^2 \tag{14}$$

Figure 12 provides four examples of distributions of response times for different values of $s$, $F$, and $t_0$.

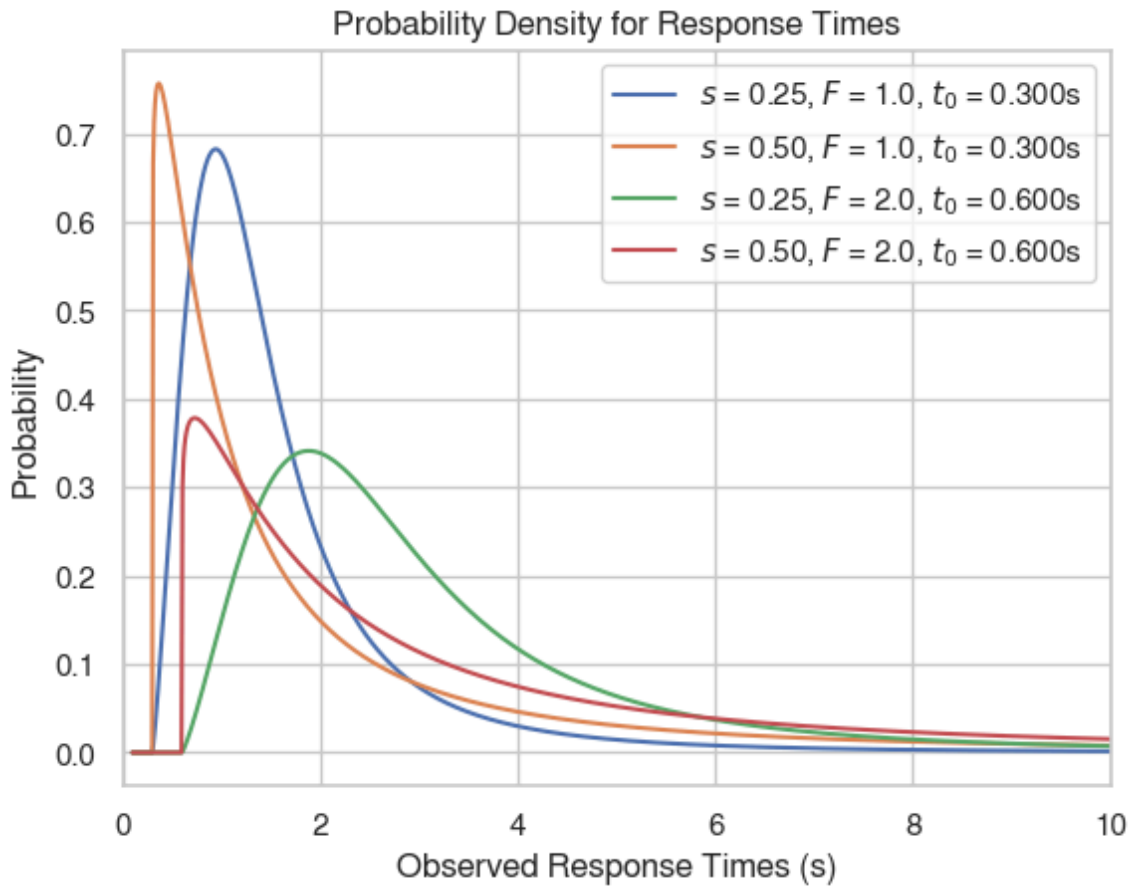

**Figure 12**: Different RT probability distributions for different values of parameters $s$, $F$, and $t_0$, assuming the retrieval of a chunk with activation $A(c) = 0.0$.

Using the equations above, we can now calculate the trial-by-trial likelihoods. One of the main advantages of this formulation is that it saves a significant amount of processing time, relative to Monte Carlo simulation. The correct likelihood values are collected in a single pass instead of hundreds or thousands of forward simulations. This makes it possible, for example, to obtain much finer-grained grid-searches and much more stable convergence when using optimization algorithms. Figure 13 shows the trial-by-trial log-likelihood of the model for

Participant 1, together with the path followed by Powell's optimization method and the combination of parameters that is most likely to fit the data.

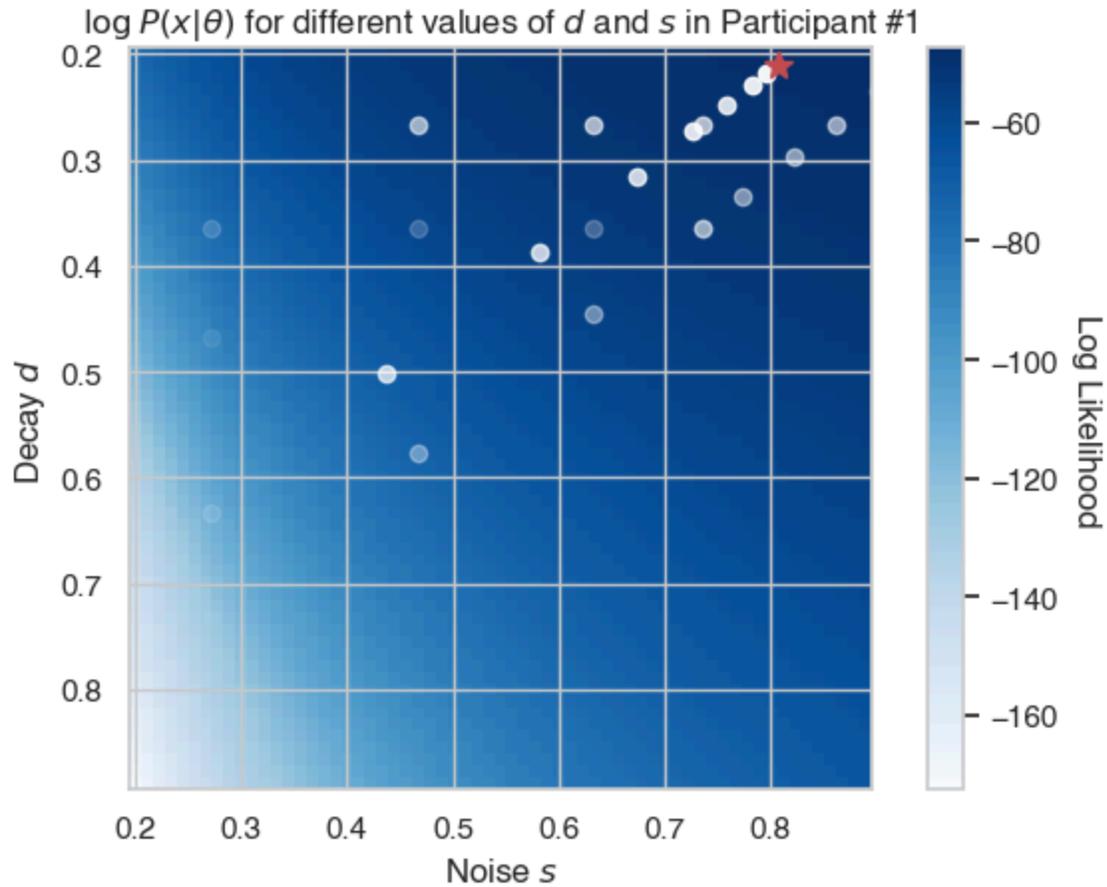

**Figure 13**: Distribution of log-likelihood values across the parameter space for

Participant 1. Points represent the path examined by Powell's method; the red star represents the final value.

**Relationship between trial-by-trial and aggregate measures**

The trial-by-trial log-likelihood space in Figure 13 looks very similar to that aggregated likelihood space of Figure 9; in fact, the optimization procedure converges on similar parameter values. This is perhaps not surprising, given that the aggregate measure is a function of the single trials, but it is not generally the case. In fact, more often than not, the parameters that best fit an individual trial-by-trial would be different than those that best fit an individual at the aggregate level.

In general, the choice of which procedure is better is highly dependent on the context. As noted above, in some cases, the trial-by-trial likelihood procedure might be impractical. However, there are reasons why, whenever possible, it should be used. Perhaps the most important is that a trial-by-trial analysis provides a more detailed view of a participant's behavior, while aggregated behavioral measures might obscure some subtle response patterns. For example, Participant 2 was deemed improbable to fit (Figure 7) given the strong stay probability after a win (Figure 10). But is their behavior truly *that* unlikely? After all, the task consists of only 64 trials, less than half of which will result in a reward. A high stay probability could result by chance, especially when one considers the long string of cases in which responding "more" has resulted in a win. In fact, running Powell's optimization algorithm shows that the most likely parameter values for Participant 2 are $d = 0.263$ and $s = 0.432$ (Figure 14). And these parameters result, in fact, in higher log-likelihood values than those obtained by applying the same procedure for Participant 1.

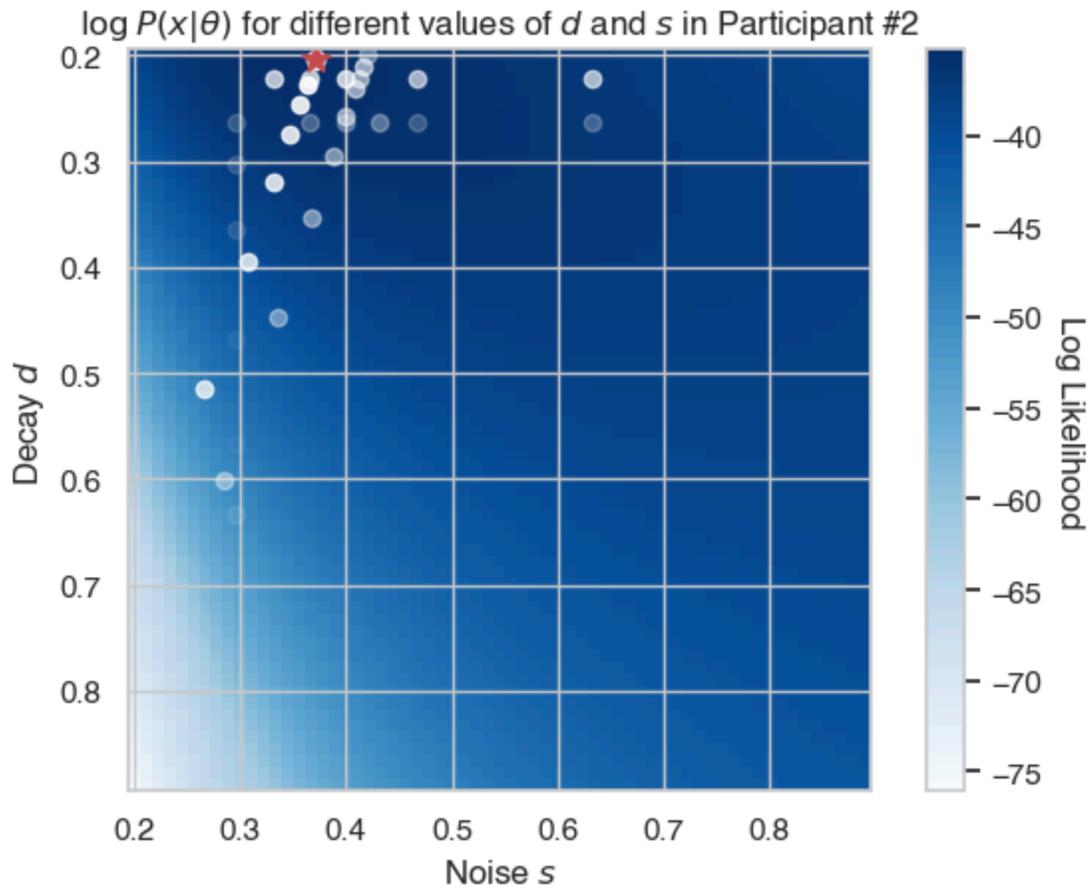

**Figure 14**: Trial-by-trial log-likelihood estimates for Participant 2's data across different values of *d* and *s*. The white dots represent the path of Powell's optimization method; the red star represents the convergence point.

## Maximum Likelihood Estimation for Group Studies

In the previous cases, we have assumed that the model is fit to data from a single participant. How does that translate to group studies with multiple participants?

Most ACT-R research papers have historically focused on fitting a model to aggregate data that averages across multiple participants. This is possible and can be done with likelihood measures as well. However, such an approach hides the true heterogeneity of the data: Group averages might not reflect the performance of any of its constituent participants. Yang, Karmol, and Stocco [54], for example, compared three different ACT-R models of syntactic parsing and found that, while data at the group level were best fit by one model, data at the individual level were consistently better fit by an alternative model, and the apparent fit of the former was due to an artifact of averaging.

Fitting different individuals, however, might give inconsistent results. Some individuals might be better fit by one model while others might be better fit by a competing model. So, how do we translate individual-level log-likelihood values into group-level values? A naive but useful assumption [55] is that the individuals in a group are independent observations or, equivalently, that fitting a model is a "fixed-effects"' process. Under these assumptions, the group likelihood $\mathscr{L}_G$ that a model would fit $N$ individuals in a group is the product of the probabilities that the model would fit each of them:

$$\mathscr{L}_G = \prod_i^N \mathscr{S}(\mathbf{x}_i \mid m, \boldsymbol{\theta}_i) = \prod_i^N P(\mathbf{x}_i \mid m, \boldsymbol{\theta}_i)$$

And

$$\log \mathscr{L}_G = \log \prod_i^N P(\mathbf{x}_i \mid m, \boldsymbol{\theta}_i) = \sum_i \log P(\mathbf{x}_i \mid m, \boldsymbol{\theta}_i) \tag{15}$$

In practice, Equation 5 means that the log-likelihood is first calculated for each participant, and then the individual log-likelihoods are simply summed together. For example, the model described above has a maximum likelihood value of –45.88 for participant 1 and –35.64 for participant 2. Thus, at the group level, its log-likelihood value is log $\mathscr{L}_G$ = –45.88 + –35.64 = –81.52. Examples of this approach can be found in studies by Leonard, Hake, and Stocco [56] and Yang et al. [54].

Notice that, in Eq. 5, the parameters $\boldsymbol{\theta}_i$ might be different for any participant (hence the use of the subscript $i$). What remains the same is the model $m$; however, different individuals might be fit by different parameterizations of the same model. This is, in fact, the very reason why participants are assumed to be independent: the parameters that work best for one participant have no bearing to what works best for a different one.

### Using Likelihood for Model Selection

The possibility of calculating likelihoods sets the stage for a more formal approach to model comparison and selection, using established tools to compute exactly *how much* a model is better than others [57,58].

Consider, for example, the case of two models, $m_1$ and $m_2$. The probability that $m_1$ explains the data better than $m_2$ can be estimated using Bayes theorem:

$$
\begin{aligned}
P(m_1 \mid \mathbf{x}) &= P(\mathbf{x} \mid m_1)\, P(m_1)\, /\, P(\mathbf{x}) \\
&= P(\mathbf{x} \mid m_1)\, P(m_1)\, /\, [P(\mathbf{x} \mid m_1)\, P(m_1) + P(\mathbf{x} \mid m_2)\, P(m_2)]
\end{aligned} \tag{16}
$$

If both models are equiprobable, then $P(m_1) = P(m_2)$, and Equation 16 can be further simplified as:

$$P(m_1 \mid \mathbf{x}) = P(\mathbf{x} \mid m_1) \ / \ [P(\mathbf{x} \mid m_1) + P(\mathbf{x} \mid m_2) \ ]$$

$$= \mathscr{L}(m_1) \ / \ [ \ \mathscr{L}(m_1) + \mathscr{L}(m_2)]$$

Another common way to compare models is through *Bayes Factors* or *BF*, that is, the ratio between the two models' posterior probabilities [57,59]. Like Bayes theorem, *BF*s can also be expressed in terms of likelihoods:

$$BF_{1,2} = P(m_1 \mid \mathbf{x}) \ / \ P(m_2 \mid \mathbf{x})$$

$$= P(\mathbf{x} \mid m_1) \ / \ P(\mathbf{x} \mid m_1)$$

$$= \mathscr{L}(m_1) \ / \ \mathscr{L}(m_2)$$

And, in terms of log-likelihood:

$$\log BF_{1,2} = \log [ \ \mathscr{L}(m_1) \ / \ \mathscr{L}(m_2) \ ]$$

$$= \log \mathscr{L}(m_1) \ - \ \log \mathscr{L}(m_2) \tag{17}$$

So, Equation 17 ensures that the difference in log-likelihoods between two models corresponds to the log of the Bayes factor of the evidence in favor of the first model over the second.

There is no straightforward relation between Bayes Factors and the *p*-values used in null-hypothesis testing; not only do the two concepts come from different statistical frameworks, but an alternative model is *not* a null hypothesis. For this reason, it is more common to use agreed-upon scales to translate *BF* values into corresponding amounts of evidence. Perhaps the most common is the one suggested by Kass and Raftery [59], who suggested that the evidence for one hypothesis or model over the alternative is "positive" if the corresponding *BF* is between 0 and 20; "strong" if the *BF* is between 20 and 150, and "decisive" if the *BF* is > 150.

**Is My Model Better Than Chance?**

One of the advantages of using likelihood is that it is straightforward to use Bayes' theorem or Bayes Factors to answer a question that often haunts researchers: is the model better than *chance*? In case of aggregate data, it is easy to create and simulate models that generate random responses. If we consider accuracies in a trial-by-trial analysis, a model that responds randomly has a constant probability ($1/N_R$) of producing a given response, where $N_R$ is the number of possible responses. The log-likelihood of such a model is, therefore:

$$\log \mathscr{L}(m_{\text{random}}) = N_T \log (1/N_R)$$

where $N_T$ is the total number of trials in the data. Because the IPT has $N_T = 64$ trials and $N = 2$ possible responses, a random model has a log likelihood of –44.36. Does our declarative model do better than chance?

To answer this question, let us consider the case of Participant 2. With best parameter values $\theta*$ (i.e., $d = 0.263$, $s = 0.432$; Figure 14), the highest log-likelihood value for the model was −35.65. According to Equation 16, the probability the model is better than a chance model is then $e^{-35.65} / (e^{-35.65} + e^{-44.36}) = 0.998$. This is confirmed by the use of Bayes factors. Using Equation 17 and, once more, exponentiating the terms to remove the logs, we find that the Bayes factor for the declarative model over the chance model is $e^{-35.65 + 44.36} = 6086.09$, which indicates "decisive" evidence in favor of the declarative model according to Kass and Raferty [59].

### An Example: Procedural vs. Declarative Decision-Making

In most realistic cases, however, a researcher would have two alternative, plausible models to compare. To examine this scenario, we will begin referring to the previously described model as the *declarative model* and introduce a second model—one that takes a complementary approach to decision-making and uses procedural rules (the other form of long-term memory in ACT-R) to make decisions. We will refer to this second model as the *procedural model*.

In the procedural model, the two possible choices for "more" and "less" are represented by two competing production rules. Instead of memorizing the result of each decision ("Reward" or "Punishment", the procedural model encodes the corresponding monetary outcome as a numerical value ($1.0 for "Win" and −$0.5 for "Loss"), which is then passed to the reinforcement learning of Equation 3 as as the value $r_t$. This value is used, in turn, to update the utility of the production rule that was used to decide. As they accrue higher utilities, productions become more likely to fire.

Like the declarative model, the procedural model has two parameters only, the learning rate α and the utility noise τ. As in the case of the declarative model, we can quickly explore the likelihood across the parameter space and use Powell's optimization algorithm to find the best fitting parameters. The results are shown in Figure 15.

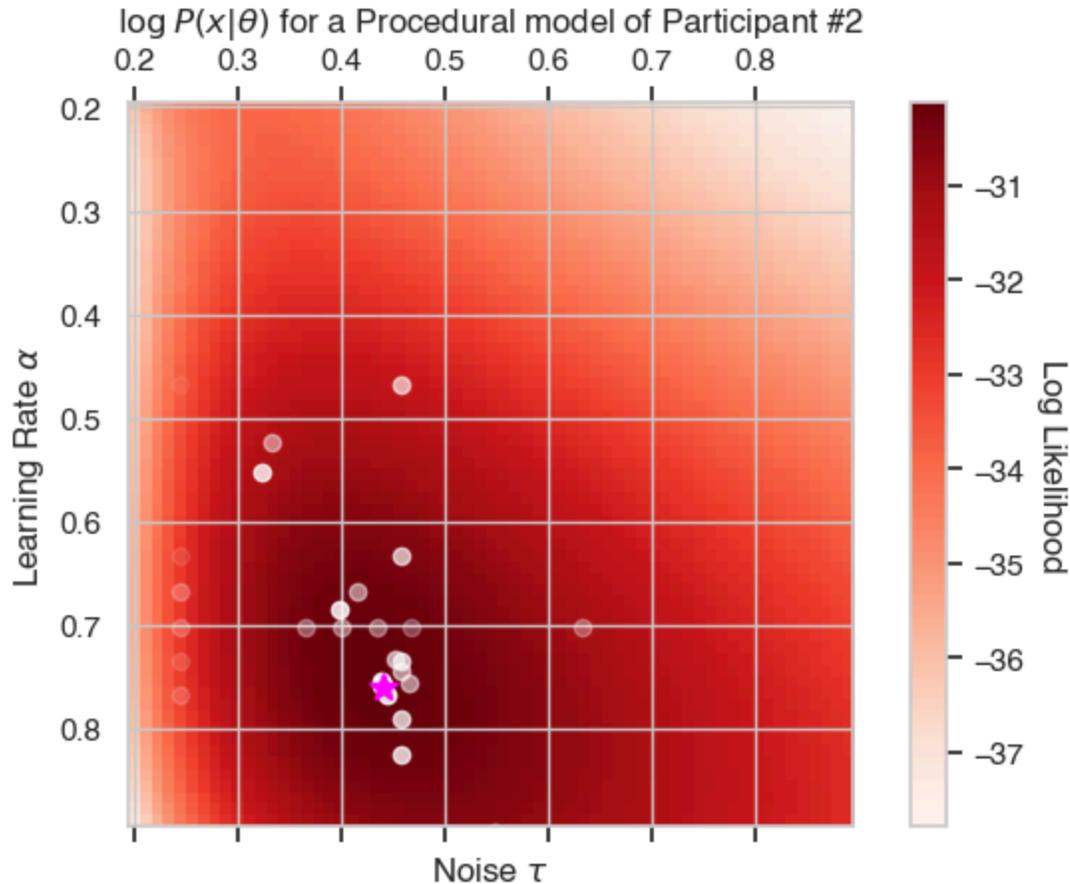

**Figure 15.** The distribution of log-likelihood values for different values of the two parameters α and τ in the procedural model. White dots represent the path of Powell's optimization algorithm; the magenta star represents the convergence point.

A comparison between Figures 14 and 15 also shows that, in general, the log-likelihood values of the procedural model are almost always higher than those of the declarative model. For the best-fitting parameters **θ\*** of the declarative model ($d$ = 0.263, $s$ = 0.432), the highest log-likelihood value was −35.65. For the best fitting parameters of the procedural model (α =

0.759 and $\tau = 0.440$, Figure 15), the highest log-likelihood value is $-30.11$. Is this a significant difference?

As in the case of the comparisons to the random model, we can answer this question by plugging in the likelihood values (after removing the logs) in Equations 16 and 17. The probability that the procedural model is then $e^{-30.11} / (e^{-30.11} + e^{-35.65}) = 0.996$, suggesting that the procedural model is almost certainly the more likely model for the data. This is confirmed by the use of Bayes factors: after exponentiating the terms to remove the logs, we find that the Bayes factor for the procedural model over the declarative model is $e^{-30.11 + 35.65} = 254.68$, which is again in the "decisive" category of evidence in favor of the procedural model according to Kass and Raferty [59].

**Comparing Models of Different Complexity**

Both the declarative and procedural models had two parameters. This makes their comparison simple, as there are no other degrees of freedom that can make one model fit the data better than the other one. In most cases, however, the researcher is not going to be so lucky; more likely than not, the comparison will involve models that contain a different number of parameters. Therefore, when evaluating a model, its likelihood should somehow be balanced by its complexity.

As noted in the introduction, one of the advantages of using likelihood is that the most common metrics that exist to balance fit and complexity, namely the Akaike Information Criterion (AIC: [28]) and the Bayesian Information Criterion (BIC: [29]), are explicitly defined in terms of likelihood. Specifically:

$$\text{AIC} = 2k - 2 \log \mathscr{L} \tag{18}$$

$$\text{BIC} = k \log n - 2 \log \mathscr{L} \tag{19}$$

where $k$ is the number of parameters in the model and $n$ is the number of observations in the data the model is fit to (i.e., the length of $\mathbf{x}_o$). Notice that, because the log likelihood term has a negative sign in Equations 18 and 19, AIC and BIC scores need to be interpreted in the opposite direction as log-likelihood values—that is, *lower* values indicate greater fit.

To see an example of the use of these metrics, let us consider a second type of declarative model. In the traditional ACT-R formulation (Equation 4) all traces share the same decay rate, $d$. Pavlik and Anderson [21,60,61], however, found that a trace's decay should be related to the overall memory activation to account for the spacing effect. Specifically, they proposed that the decay rate $d(i)$ of the $i$-th trace should be

$$d(i) = \beta e^{A(c)} + \varphi \tag{20}$$

Where $\beta$ is a parameter that modulates the spacing effect and $\varphi$ is an individual's characteristic rate of forgetting [19–21,62]. Unlike the previous declarative model, which had two parameters ($d$ and $s$), this model has three parameters ($\beta$, $\varphi$, and $s$). Running Powell's method on this model finds that, for Participant 1, the best combination of parameter values ($\beta = 0.87$, $\varphi = 0.10$, and $s = 0.89$) yields a log-likelihood value of $-43.32$, which is, in fact, better than the corresponding value ($\log \mathscr{L} = -45.88$) for the two-parameter model (with $d = 0.43$, $s = 0.87$). While the difference is not particularly large, it still corresponds to a Bayes factor of 12, which is "moderate" according to Kass's guidelines. But how much of the improvement is actually offset by the additional complexity?

Table 2 compares the two models in terms of both raw log-likelihood and the AIC and BIC metrics, with the best-fitting model for each measure highlighted in boldface. As the table shows, the three-parameter model does provide an objectively better fit even when its additional complexity is accounted for.

**Table 2**: A comparison of the two declarative models in terms of log-likelihood, AIC (Equation 18) and BIC (Equation 19).

| Criterion | Two-parameter model ($k = 2$, $n = 64$) | Three-parameter model ($k = 3$, $n = 64$) |
|---|---|---|
| Log-likelihood | –45.88 | **–43.32** |
| AIC | 95.76 | **92.64** |
| BIC | 100.07 | **99.12** |

**Generalization and the Bias-Variance Trade-off in Maximum Likelihood Estimation**

In any statistical learning problem, there is an inherent trade-off between *bias* (error due to erroneous or overly simplistic assumptions in the learning algorithm) and *variance* (error due to excessive complexity in the learning algorithm allowing it to capture random noise in the data). High bias can lead to *underfitting*, where the model is too simple to capture the complexity of the data. On the other hand, high variance often leads to *overfitting*, where the model (usually over-parameterized) captures random noise instead of the underlying data trend. Achieving a good balance between bias and variance is essential for optimal model performance and *generalization*, which is the model's ability to perform well on new, unseen data. This trade-off

is quantitatively addressed using the MSE of an estimator $\hat{f}$ which decomposes into:

$$MSE(\hat{f}) = Bias(\hat{f})^2 + Variance(\hat{f}) \tag{21}$$

This decomposition illustrates how both bias and variance contribute to the total error. In the context of Maximum Likelihood Estimation (MLE), one can show that MLE is asymptotically unbiased, which means that as the sample size increases, the bias of the MLE tends toward zero.

It is important to note, however, that MLE may still exhibit significant bias for finite samples,

particularly if the model is complex or the data are sparse. The variance of MLE decreases as the

sample size increases, reflecting the notion that more data provides more information, leading to

more precise estimates. Methods such as L1 [63] and L2 [64] regularization manage the

bias-variance trade-off by penalizing complexity during the optimization process and the

constraints can be incorporated in prior distributions over the model parameters in a Bayesian

setting (See below, "Maximum a Posteriori Estimation" section). For model selection, AIC and

BIC provide criteria to balance fit and complexity, as discussed above.

## Special Cases

### Comparison of Nested Models

A special case of model comparison is the case in which one model is *nested* within the

other. A model $m_1$ is said to be nested within $m_2$ if it contains all of the parameters of $m_2$, plus

additional parameters. Since it contains additional parameters, $m_1$ can be seen as a specialization

of $m_2$, and $m_2$ as a more general version of $m_1$. An interesting example of this case is offered in

Haile et al. (2020), which compares three different decision-making models: one RL model that relies on procedural reinforcement learning ($m_{RL}$), one declarative memory model that relies only on long-term memory ($m_{LTM}$), and one mixed model that dynamically switches between the two ($m_{mixed}$). The mixed model inherits all the parameters of the RL model and the LTM model and is, therefore, nested under both of them.

When one model is nested under another, it is possible to assign a *p*-value to the difference in log-likelihoods between two models using Wilks' theorem [65]. The theorem provides a way to interpret their difference in likelihood under a null hypothesis testing framework, and automatically accounts for the different complexity of the two models. If we define $\Delta \mathscr{L} = \log \mathscr{L}(m_1) - \log \mathscr{L}(m_2)$, and the two models contain $k_1$ and $k_2$ parameters, respectively, then Wilks' theorem states that the probability that their difference in likelihoods is greater than what would be expected by chance is approximated by the area beyond $2\Delta \mathscr{L}$ in a $\chi^2$ distribution whose degrees of freedom are the difference in parameters $\Delta k = k_1 - k_2$.

In our case, it is possible to see the three-parameter memory model as nested within the standard memory model. This is because, when $\beta = 0$, Equation 20 reduces to a constant term $d = \varphi$ for all traces, as in Equation 4. In fact, it can be shown that, when $\beta = 0$, the parameter space for $\varphi$ and s is identical to that of $d$ and $s$. Thus, we can apply Wilks' theorem, using $\Delta \mathscr{L} = -43.32 + 45.88 = 2.56$ and $\Delta k = 3 - 2 = 1$. Figure 16 provides a visual rendition of the results, with the blue line representing the $\chi^2$ distribution and the orange line representing $\Delta \mathscr{L}$. The small portion of the area under the curve beyond the orange line, highlighted in solid orange, represents the corresponding *p*-value of the difference in fit between the two models. In this case, $p = 0.02$, further confirming that the additional gain in fit more than compensates for the additional degree of freedom.

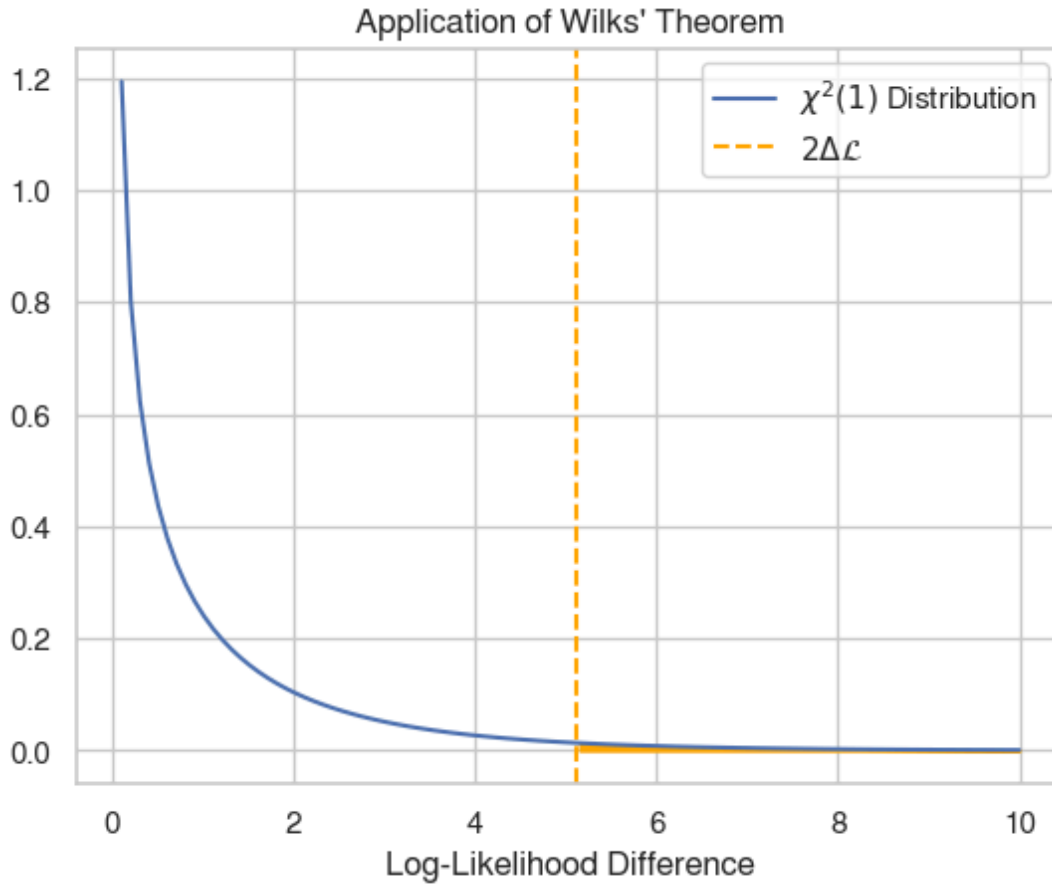

**Figure 16**: Wilks' theorem applied to the difference between the 2-parameter and the 3-parameter versions of the declarative model.

## Maximum *A Posteriori* Estimates

In some cases, reasonable constraints can be posed for the value of parameters. Some researchers, including some of the authors of this manuscript, have argued that ACT-R parameters have a biological interpretation: they reflect neurophysiological quantities such as the density of dopamine receptors or the connectivity between different brain areas [19,37,38] . If this is true, it would be reasonable to expect that the values of these parameters are not uniformly distributed. The problem, in general, is to find the ideal set of values that maximize the posterior probability of a set of parameters $\boldsymbol{\theta}$, given the data, $\mathbf{x}$, i.e., argmax$_\theta$ $P(\boldsymbol{\theta} \mid \mathbf{x})$. From Bayes theorem, we know that the posterior probability $P(\boldsymbol{\theta} \mid \mathbf{x})$ can be expressed as a product of the likelihood and the prior:

$$P(\boldsymbol{\theta} \mid \mathbf{x}) = P(\mathbf{x} \mid \boldsymbol{\theta})\, P(\boldsymbol{\theta}) \,/\, P(\mathbf{x}) \qquad (21)$$

The term $P(\mathbf{x})$ in Equation 21 can be ignored: given that $\mathbf{x}$ is the target data whose probability we are trying to maximize, it represents just a constant term that does not affect which parameters are the most likely. And, conceptually, we are not concerned with what the probability of the observed data is—no matter what it is, we are trying to replicate it.

Based on Equation 21, we can express the most likely parameters $\boldsymbol{\theta}^*$ as:

$$\text{argmax}_\theta\, P(\boldsymbol{\theta} \mid \mathbf{x}) = \text{argmax}_\theta\, P(\mathbf{x} \mid \boldsymbol{\theta})\, P(\boldsymbol{\theta}) \; = \text{argmax}_\theta\, \mathscr{L}(\boldsymbol{\theta} \mid \mathbf{x})\, P(\boldsymbol{\theta}) \qquad (22)$$

In other words, a more precise estimate of the best parameters $\boldsymbol{\theta}^*$ can be obtained by maximizing the product of the likelihood function $\mathscr{L}(\boldsymbol{\theta} \mid \mathbf{x})$ and the prior probability of the parameters $P(\boldsymbol{\theta})$. Because it maximizes the posterior probability of $\boldsymbol{\theta}$, this approach is known as a maximum *a posteriori* estimation or MAP.

In most cases, the MAP approach is not useful because the probability distributions of the ACT-R parameter values are unknown. ACT-R provides mean default values for many parameters, but no information about their distribution. In some cases, however, these distributions can be reasonably approximated. Xu and Stocco [17], for example, used the model developed by [38] to determine the most likely values of two procedural learning parameters for each participant performing a 2AFC task.

For at least one ACT-R parameter, however, researchers have been able to empirically measure a distribution of values: the $\varphi$ parameter in Equation 20. Since its introduction in Pavlik and Anderson [60], numerous studies have reported consistent values and distributions across different studies in which the equation was fit to memory data for individual participants [21,62]. For example, Zhou et al [19] reported that, across 50 young adults, the values of phi for verbal stimuli were approximately normally distributed with a mean of $\varphi = 0.283$ and SD = 0.062. Using these values as a reference, we can calculate the prior probability of the best-fitting parameter $\varphi = 0.01$ obtained from Powell's method in the previous section: $P(\varphi = 0.01) = 0.00002$. Such a value is, therefore, *extremely* unlikely—even if it allows the model to fit the data (Figure 17).

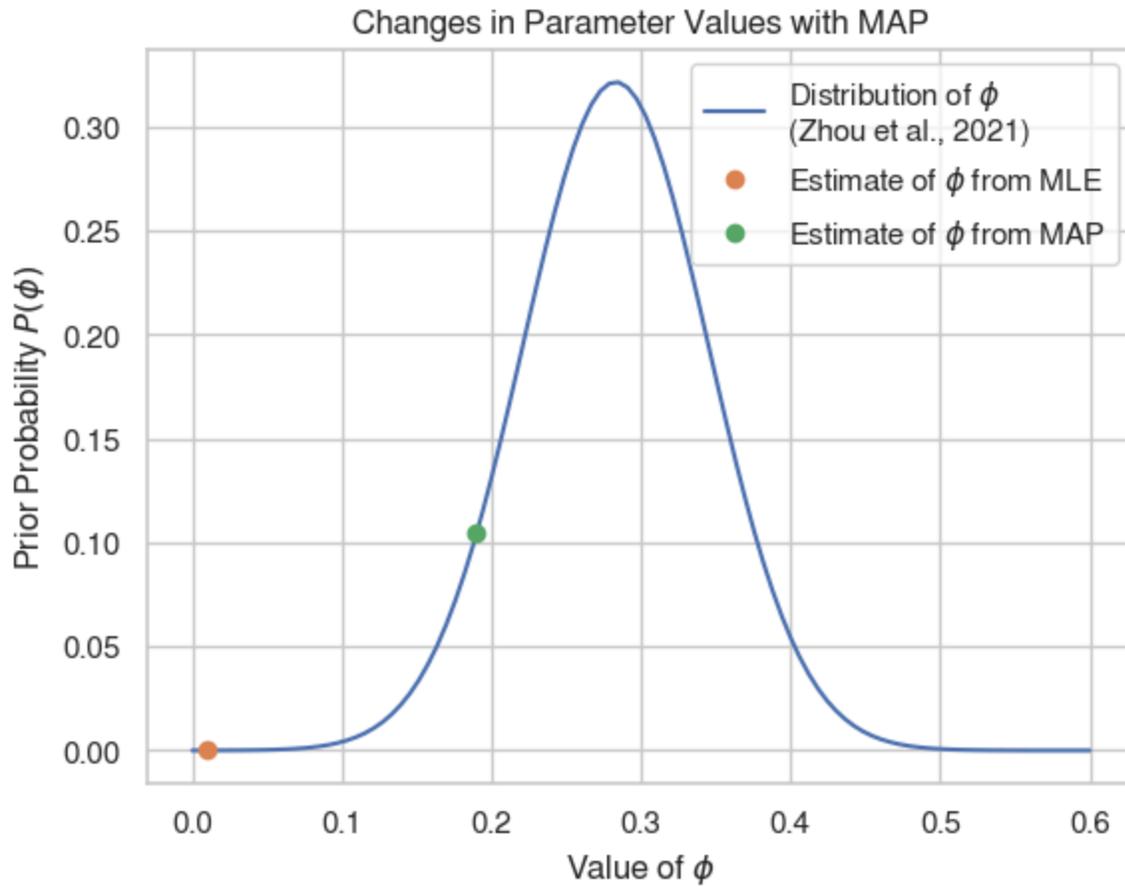

**Figure 17**: Distribution of the φ parameter in the data of Zhou et al (2021; blue line). When the distribution is used as part of the MAP procedure, the best-fitting value changes from φ = 0.01 (orange point) to φ = 0.19 (green point).

To obviate this problem, we can re-run the model-fitting procedure so that it uses Powell's method to maximize the posterior value instead of the likelihood. As before, we can use log values to avoid vanishingly small probabilities, thus maximizing the quantity:

$$\text{argmax}_{\theta} \log(\ \text{argmax}_{\theta}\ \mathscr{L}(\boldsymbol{\theta} \mid \mathbf{x})\ P(\boldsymbol{\theta})\ ) =\ \text{argmax}_{\theta}\ (\log \mathscr{L}(\boldsymbol{\theta} \mid \mathbf{x}) + \log P(\boldsymbol{\theta})\ )$$

One theoretical problem is that $P(\boldsymbol{\theta})$ is, once more, a *joint* probability distribution of parameters. The model in question has three parameters ($\varphi$, $\beta$, and $s$) and, while we know the distribution of $\varphi$, we do not know about $\beta$ and $s$, nor do we know their correlations. Lacking this information, we need to make some assumptions before applying the MAP procedure. Specifically, we will assume that (1) the three parameters are uncorrelated with one another; and (2) that $\beta$ and $s$ are uniformly distributed. The first assumption implies that $\log P(\boldsymbol{\theta}) = \log P(\varphi)P(\beta)P(s) = \log P(\varphi) + \log P(\beta) + \log P(s)$. The second assumption implies that $P(\beta)$ and $P(s)$ are constant and, therefore, can be ignored when maximizing the posterior probability (as their values do not change when the corresponding parameter value changes). Thus, when running the MAP procedure, we can approximate $\log P(\boldsymbol{\theta})$ as $\log P(\varphi)$.

When applied, the MAP procedure yields a different combination of parameter values, now converging on $\varphi = 0.19$, $\beta = 0.04$, and $s = 0.86$. As expected, this procedure balances the fit of the data with the reasonableness of the parameter and yields a much more probable value of $\varphi$: $P(\varphi = 0.19\ ) = 0.10$ (Figure 17). Notice that, as a consequence of this additional constraint posed on $\varphi$, the value of $\beta$ has dramatically changed as well.

**Discussion**

In this paper, we argued that likelihood should be used in fitting, evaluating, and comparing cognitive architecture models and provided a tutorial with several worked-out examples of how this could be achieved in ACT-R. Using likelihood offers several advantages, including a transparent interpretation of its values, the possibility of combining data on different scales (e.g., response times and accuracies), the versatility in handling both single-subject and group-level analysis, and, ultimately, the benefits that come from adopting a measure that is well-established and vastly used in statistical analysis across diverse scientific and technical domains.

Because of its nature, this primer is limited to relatively simple examples. All of the cases illustrated come from simple decision-making tasks; in these tasks, human behavior can be modeled with the selection of a single production rule or the retrieval of a single memory. However, most applications of cognitive architectures contain a mixture of different mechanisms; in ACT-R, even simple tasks often contain a mixture of different productions and memory retrievals. For example, existing models of response interference paradigms [66,67] rely on the serial application of different chains of production rules interleaved with different memory retrievals. While we have not provided examples with similar models, the methods presented herein would naturally scale up in at least two ways. Although there are times when it is computationally impractical at scale, it is *always possible* to run multiple forward simulations of the model and compute maximum likelihood over the aggregated data, no matter how complex the model is. Second, it is possible to compute the exact trial-by-trial probabilities associated with a single response and response time even if the model includes mixtures of production rules and declarative retrievals. In the latter case, one would need to compute the probabilities associated with all of the possible paths (i.e., series of production rules) that lead to a given response. Explicitly laying out the paths of production firings and declarative retrievals that lead to a response might be onerous, but is not impossible—it is the foundation, for example, of the analysis technique known as Multinomial Processing Trees [68].

Despite its popularity, maximum likelihood estimation is not without problems. In model selection, for example, maximum likelihood *per se* is not a good metric. BIC and AIC, in fact, have been developed as ways to handle the shortcomings of maximum likelihood, and, as we have shown, both can be used for architecture models. Some authors have argued for other model selection methods that are not based on Maximum Likelihood, such as parameter space partitioning [PSP: 69] and model flexibility analysis [MFA: 42]. Both alternatives provide important advantages. Most notably, because they are solely based on the analysis of the possible model outputs independently of its parameters, they can be used to compare vastly different models—for example, an ACT-R model against a neural network model. Still, it is not clear how to quantify model differences in PSP beyond a qualitative analysis of their parameter space, and it has been argued that MFA is functionally equivalent to Normalized Maximum Likelihood [70] — which is, of course, related to likelihood itself.

As noted in the introduction, one important caveat is that likelihood is designed for *parametric* models, but cognitive architectures contain both parametric and non-parametric aspects. For example, in addition to scalar quantities and parametric processes, ACT-R models also contain discrete, symbolic chunks and productions, which are non-parametric in nature. The characteristics of these structures affect how the models work, but they are transparent to the likelihood and, because they are not associated with parameters, are not included in AIC and BIC measures. Because of this, the real complexity of cognitive architecture models is likely higher than what can be glimpsed by the number of free parameters [71].

These limitations notwithstanding, we believe that using likelihood measures, owing to their versatility and transparency, would be beneficial for researchers using cognitive architectures. We hope this paper facilitates its adoption in the future.

## Acknowledgments

This work was partially supported by a grant from the Garvey Institute for Brain Health Solutions to AS. Participation in the preparation of this manuscript by KM and KG was supported by a combination of Grant FA9550-23-1-0661 from the Air Force Office of Scientific Research, the Assessing and Augmenting Performance in Extreme Environments (A$^2$PEX) Cooperative Agreement with the Air Force Research Laboratory. The content of this manuscript does not necessarily reflect the views of the USAF, AFRL, or AFOSR.

129–134.